\documentstyle[pra,aps,preprint]{revtex}
\begin{document}
\draft
\title{Theory of Quantum Electrodynamical Self-consistent Fields \footnote{This work is firstly presented at the meeting of the physical society of Japan at Iwate univ., Sep. 1999, and given as a part of the doctoral thesis of the author presented to Osaka univ., Nov. 2000.}}
\author{Tadafumi Ohsaku}
\address{Department of Physics, Graduate School of Science, Osaka University, Toyonaka-shi Machikaneyama-machi 1-1 Osaka, 560-0043, Japan}

\maketitle

\begin{abstract}
To obtain the basis for combining various many-body techniques to QED in a consistent manner, we investigate the theory of quantum electrodynamical self-consistent fields.
The reserch interest was born mainly of the electronic structure theory, thus we consider of atomic and moleculer systems as our main subjects. But the formalism is more fundamental.
First, we derive the quantum electrodynamical Hartree-Fock theory by using the Green's function method. Then we construct a relativistic
 Hamiltonian written by creation-annihilation operators for electron and positron in a general form, and check that it reproduces the Hartree-Fock theory. We use this Hamiltonian to derive the time-dependent Hartree-Fock theory and 
random  phase approximation, in the operator formalism. The relativistic Slater determinant of the Thouless parametrization is also used.
Finally we discuss the applications and futher possible developments. 
\end{abstract}

\section{Introduction}
\label{sec:intro}

It is natural to have an interest in 
the method to construct the relativistic many-body theory ( and relativistic 
self-consistent fields ). There are also many physical situations demanding 
the relativistic many-body theory for atoms, molecules and solids.
Relativistic effects appear to be crucial in many cases. 
We have several relativistic extensions of many-body theory\cite{1,2,3}, 
but almost all of them are approximate theories from the beginning, 
not the fully relativistic formalism. The Hamiltonian is given under some assumption and approximation. Therefore they have validities and limitations.
In the case of relativistic electronic structure theory, there are several approximation schemes. The famous Dirac-Fock method is based on the Dirac-Coulomb-Breit no-sea (the approximation completely neglects the effects of the Dirac sea) Hamiltonian\cite{1}. It has a Dirac type one-body Hamiltonian, and 
describes the interaction of electrons via the Coulomb potential and Breit operator. This Hamiltonian with several many-body schemes (Hartree-Fock, relativistic many-body perturbation theory, so on) give good numerical results for neutral heavy elements. But it can not be applied, for example, to describe the inner-core electrons of heavy elements or, electrons of highly ionized heavy elements such as lithium-like uranium. Because in these cases, quantum electrodynamical, QED, effects (such as the vacuum polarization or the electron self-energy) can not be negrected, we must take the Dirac sea into account.
There is also a theoretical treatment for taking account of these QED effects\cite{3}. This method is based on the Furry picture of QED, and the QED corrections are estimated perturbatively with the renormalization procedure. This method gives very accurate results for highly charged ions. But it gives only correction. This method is not based on a single principle from beginning to the end: 
It must be combined with the Dirac-Coulomb-Breit no-sea scheme, 
for comparing with the experimantal value.  
Practically it treats only the one-particle QED effects, 
and the problem is reduced
to that of hydrogen-like atoms. A matter of course, this method is only 
valid for the systems of a few electrons.

A theory suitable for description of the interaction between electrons 
is intrinsically QED\cite{4}. Electrons interact by exchaging a photon 
intermediately. 
Only QED can fully describe such a situation. 
Thus we have a fundamental question that how to construct a 
relativistic many-body theory based on QED in a consistent manner. 
And also how we can construct the self-consistent fields in QED, 
which it should be the central theme of the many-body theory in QED. 
In the nonrelativistic theory, there are various many-body techniques, 
and these methods are used in atoms, molecules, solids and nuclear systems. 
A large part of them is constructed in the operator method, 
and people use them in discussions, not only for stational properties 
but also for dynamical processes.
It should be fruitful if these many-body techniques are applied to QED.
It should enlarge the ability of QED, and should overcome the limitation
of validity of it. 
By these considerations, we investigate a general formalism 
for the many-body theory in QED, especially the self-consistent mean-fields. 
In Sec.\ref{sec:der}, we derive the quantum electrodynamical Hartree-Fock (HF) 
theory by using the Green's function formalism. 
By this method, we can introduce the HF theory with no ambiguity. 
Thus we regard the result as the basis for our next step of investigation. 
In Sec.\ref{sec:ope}, we give a relativistic Hamiltonian which is explicitly 
written by the creation-annihilation operators. 
We present a evidence that this 
Hamiltonian obtains the same result for the HF, with the Green's function 
method given in Sec.\ref{sec:der}. With this Hamiltonian, 
we examine the method introduced from the nonrelativistic many-body theory 
in the operator formalism, and derive the time-dependent Hartree-Fock (TDHF) theory
, HF condition, and random phase approximation (RPA). 
We expect that the operator formalism will extract the capability of our 
theory. In this paper we mainly focus on the formalism. But we also discuss 
the possibilities of our theory, especially in applications of atomic physics,
 in Sec.\ref{sec:app}. 
In Sec.\ref{sec:con}, some general conclusion with summary, discussions of the 
remaining problems and the futher developments, are presented. 
We also present the finite-temperature quantum electrodynamical 
HF theory in appendix.

\section{Derivation of the Hartree-Fock Theory by using Green's function formalism}
\label{sec:der}

In this section we introduce a QED Lagrangian and Hamiltonian, and then we derive the Hartree-Fock theory by the field-theoretical Green's function method.
Some parts of the contents were given in literature\cite{5,6,7,8,9}. But Eq.(21) is our original result. The form of Eq.(21) will become the basis of our theory developed in Sec.\ref{sec:ope}. Here we intend to confirm the theory and to construct the basis for later investigation. 
We use the metric convention as
$ds^{2} = g_{\mu\nu}dx^{\mu}dx^{\nu} = (dx^{0})^{2} - (dx^{1})^{2} - (dx^{2})^{2} - (dx^{3})^{2}$, and the metric tensors are defined as
$g_{\mu\nu}={\rm diag}(1,-1,-1,-1)$.

The starting point for our discussion is the following QED Lagrangian
\begin{eqnarray}
{\cal L}(x) &=& -\frac{1}{4}F_{\mu\nu}(x)F^{\mu\nu}(x) + \bar{\psi}(x)(i\gamma^{\mu}\partial_{\mu}-m_{0})\psi(x) +e\bar{\psi}(x)\gamma^{\mu}\psi(x)(A_{\mu}(x)+A^{(e)}_{\mu}(x)),
\end{eqnarray}
where $\psi(x)$ designates the electron-positron field, $m_{0}$ as its bare mass, and $\gamma_{\mu}$ is the usual Dirac gamma matrix. The convention $e=|e|$ is used. The anti-symmetric electromagnetic tensor is defined by
$F_{\mu\nu}=\partial_{\mu}A_{\nu}-\partial_{\nu}A_{\mu}$.  $A_{\mu}$ is the electromagnetic potential and $A^{(e)}_{\mu}$ is the external field. To consider 
the atomic situation, we introduce the static Coulomb potential of 
nucleus and write it as $A^{(e)}_{0}=U(r)$, where it is in fact 
nonrelativistic and 
classical. Now the coordinates describing the system are fixed at 
a nucleus which is located at the origin. As long as we treat the interaction between the nucleus and 
electrons by the external field approximation, it is difficult to describe the 
retardation effects. We suppose it is difficult to construct a fully Lorentz 
covariant theory in our situation. But it is enough for our aim in this paper.
 Here we also negrect the recoil effect of the nucleus. 

Adopting the action principle, we obtain the equations of motions (in the Lorentz gauge)
\begin{eqnarray}
\partial_{\mu}F^{\mu\nu} = (\partial^{2}_{0}-\nabla^{2})A^{\nu} &=& -e{\bar \psi}\gamma^{\nu}\psi,  \\
(i\gamma^{\mu}\partial_{\mu}-m_{0}-\gamma_{0}U(r))\psi &=& -e\gamma^{\mu}A_{\mu}\psi.
\end{eqnarray}
Now we use the well known canonical quantization scheme for the Dirac fields 
\begin{eqnarray}
\{\hat{\psi}_{\alpha}(x_{0},{\bf x}),\hat{\psi}^{\dagger}_{\beta}(x_{0},{\bf y})\} &=& \delta_{\alpha\beta}\delta({\bf x}-{\bf y}), \\
\{\hat{\psi}_{\alpha}(x_{0},{\bf x}),\hat{\psi}_{\beta}(x_{0},{\bf y})\} &=& \{\hat{\psi}^{\dagger}_{\alpha}(x_{0},{\bf x}),\hat{\psi}^{\dagger}_{\beta}(x_{0},{\bf y})\} = 0,
\end{eqnarray}
where $\alpha$ and $\beta$ denote the spinor indices. We also quantize the electromagnetic field $A_{\mu}$, but here we do not discuss the problem of quantization scheme with gauge fixing. For our aim, it is unnecessary. The equation of motion for 
photon are formally solved as
\begin{eqnarray}
\hat{A}_{\mu}(x) = -e\int d^{4}y D_{\mu\nu}(x-y)\hat{\bar{\psi}}(y)\gamma^{\nu}\hat{\psi}(y), 
\end{eqnarray}
where $D_{\mu\nu}(x-y)$ is a full 2-point function for photon. Here we assume there are only virtual photons in the system. Next, we remove the photon degree 
of freedom and the equation of motion for the Dirac field becomes
\begin{eqnarray}
i\gamma^{0}\partial_{0}\hat{\psi}(x) = (-i\gamma\cdot\nabla+m_{0}+\gamma_{0}U(r))\hat{\psi}(x) + e^{2}\int d^{4}y\gamma^{\mu}\hat{\psi}(x)D_{\mu\nu}(x-y)\hat{\bar{\psi}}(y)\gamma^{\nu}\hat{\psi}(y).
\end{eqnarray}
Then we find a Hamiltonian operator for describing the interacting many-electron-positron system
\begin{eqnarray}
\hat{H} &=& \int d^{3}x\hat{\bar{\psi}}(x)(-i\gamma\cdot\nabla+m_{0}+\gamma_{0}U(r))\hat{\psi}(x) \nonumber \\
  & & +\frac{1}{2}e^{2}\int d^{3}x\int d^{4}y \hat{\bar{\psi}}(x)\gamma^{\mu}\hat{\psi}(x)D_{\mu\nu}(x-y)\hat{\bar{\psi}}(y)\gamma^{\nu}\hat{\psi}(y). 
\end{eqnarray}

After these preparations, we derive the HF theory by using the Green's functions and Feynman rules. This method is famous in the textbook for the nonrelativistic many-body theory\cite{10}, and here we extend it straightforwardly. 
In the perturbation theory based on the Dyson equations, the HF approximation 
is defined as an infinite-order summation of specific diagrams in expanding the fermion 2-point function\cite{5,6,7,8,10}. Here we also obey this ansatz and take into account two diagrams as the lowest order 
vacuum polarization and fermion self-energy, in the central field.
Thus we obtain the next formula for the HF total energy
\begin{eqnarray}
E &=& -i\int d^{3}x (-i\gamma\cdot\nabla +m_{0} +\gamma_{0}U(r)){\rm tr}(S^{HF}_{F}(x,x)) \nonumber \\
       & & -\frac{1}{2}e^{2}\int d^{3}x \int d^{4}y D^{(0)}_{F}(x-y)_{\mu\nu}{\rm tr}(\gamma^{\mu}S^{HF}_{F}(x,x)){\rm tr}(\gamma^{\nu}S^{HF}_{F}(y,y)) \nonumber \\
       & & +\frac{1}{2}e^{2}\int d^{3}x \int d^{4}y D^{(0)}_{F}(x-y)_{\mu\nu}{\rm tr}(\gamma^{\mu}S^{HF}_{F}(x,y)\gamma^{\nu}S^{HF}_{F}(y,x)),
\end{eqnarray}
where the first line is the kinetic energy while the second and the third are the direct and the exchage HF energy, respectively. The trace is taken over the 
spinor indices. $D^{(0)}_{F}(x-y)_{\mu\nu}$ is the free photon Feynman propagator, and especially in the Feynman gauge, it is written as
\begin{eqnarray}
iD^{(0)}_{F}(x-y)_{\mu\nu}
   &=& -ig_{\mu\nu}\int \frac{d^{4}k}{(2\pi)^{4}}\frac{e^{-ik(x-y)}}{k^{2}+i\eta}.
\end{eqnarray}
$S^{HF}_{F}(x,y)$ is the one-body fermion propagator constructed by the variationally optimized HF one-body functions. The fermion field operators are expanded generally as
\begin{eqnarray}
\hat{\psi}(x) &=& \sum_{n}(\psi^{(+)}_{n}({\bf x})e^{-i\epsilon^{(+)}_{n}x_{0}}\hat{a}_{n} + \psi^{(-)}_{n}({\bf x})e^{+i\epsilon^{(-)}_{n}x_{0}}\hat{b}^{\dagger}_{n}), \\
\hat{\bar{\psi}}(x) &=& \sum_{n}(\bar{\psi}^{(+)}_{n}({\bf x})e^{+i\epsilon^{(+)}_{n}x_{0}}\hat{a}^{\dagger}_{n} + \bar{\psi}^{(-)}_{n}({\bf x})e^{-i\epsilon^{(-)}_{n}x_{0}}\hat{b}_{n}),
\end{eqnarray}
where $\hat{a}_{n}(\hat{a}^{\dagger}_{n})$ is the electron annihilation(creation) operator while $\hat{b}_{n}(\hat{b}^{\dagger}_{n})$ is the positron annihilation(creation) operator. Here $n$ indicates some quantum numbers suitable for the description of the system.
Especially in the HF theory, $\psi^{(\pm)}_{n}$ becomes the HF one-body function and $\pm\epsilon^{(\pm)}_{n}$ is its eigenvalue.    
The vacuum is determined by the relations
\begin{eqnarray}
\hat{a}_{n}|0\rangle &=& 0, \\
\hat{b}_{n}|0\rangle &=& 0. 
\end{eqnarray}
Hereby the HF fermion propagator is defined by
\begin{eqnarray}
iS^{HF}_{F}(x,y) &=& \langle \Phi_{0}|T\hat{\psi}^{HF}(x)\hat{\bar{\psi}}^{HF}(y)|\Phi_{0}\rangle \nonumber \\
     &=& \sum_{n} \psi^{(+)}_{n}({\bf x}) \bar{\psi}^{(+)}_{n}({\bf y}) e^{-i\epsilon^{(+)}_{n}(x_{0}-y_{0})} \theta(x_{0}-y_{0})\theta(\epsilon_{n}-\epsilon_{F}) \nonumber \\
     & & -\sum_{n} \psi^{(+)}_{n}({\bf x}) \bar{\psi}^{(+)}_{n}({\bf y}) e^{-i\epsilon^{(+)}_{n}(x_{0}-y_{0})} \theta(y_{0}-x_{0})\theta(\epsilon_{F}-\epsilon_{n}) \nonumber \\
     & & -\sum_{n} \psi^{(-)}_{n}({\bf x}) \bar{\psi}^{(-)}_{n}({\bf y}) e^{+i\epsilon^{(-)}_{n}(x_{0}-y_{0})} \theta(y_{0}-x_{0}),
\end{eqnarray}
where $|\Phi_{0}\rangle$ is the HF wavefunction occupied by electrons up to the Fermi level $\epsilon_{F}$ ( so called single Slater determinant ). Perform the Fourier transformation in time, it becomes
\begin{eqnarray}
S^{HF}_{F}({\bf x},{\bf y},\epsilon) &=& \sum_{n} \psi^{(+)}_{n}({\bf x})\bar{\psi}^{(+)}_{n}({\bf y}) \Biggl(\frac{\theta(\epsilon_{n}-\epsilon_{F})}{\epsilon-\epsilon^{(+)}_{n}+i\eta} + \frac{\theta(\epsilon_{F}-\epsilon_{n})}{\epsilon-\epsilon^{(+)}_{n}-i\eta} \Biggr) \nonumber \\
     & & +\sum_{n}\psi^{(-)}_{n}({\bf x})\bar{\psi}^{(-)}_{n}({\bf y})\frac{1}{\epsilon+\epsilon^{(-)}_{n}-i\eta}.
\end{eqnarray}
In (9), we also perform the Fourier transformation in time and employ the energy contour integration. For this purpose we must decide the integration path\cite{7}, so we will examine the structure of the 4-current because in (8) and (9), the interaction terms are written in the form of current-current interactions. If we choose the current operator which is normal-ordered
about the Dirac vacuum, its vacuum expectation value vanishes in the zeroth-order. 
Consider, for example when we introduce an external field generated by a nucleus and describe the system by using the Furry picture\cite{3,11}, 
the vacuum polarization is observed around the nucleus, 
so the vacuum expectation value of the current
becomes non-zero even in the zeroth-order\cite{3}. This situation becomes more clear if we concern to the fact that the zeroth-order in the Furry picture already involves an infinite order of summation in which each term is expanded by the external field. We want to include the vacuum polarization effect in our theory. For this purpose, the current generally have to be calculated by the Schwinger's expression\cite{8,12}
\begin{eqnarray}
j_{\mu}(x) &=& i\frac{e}{2}\Bigl(\lim_{y \to x(x_{0}>y_{0})} + \lim_{y \to x(x_{0}<y_{0})}\Bigr)_{(x-y)^{2}\ge 0}{\rm tr}(\gamma_{\mu}S_{F}(x,y)). 
\end{eqnarray}
The zeroth-order vacuum expectation value in the Furry picture becomes 
\begin{eqnarray}
j^{(0)}_{\mu}(x) &=& i\frac{e}{2}\Bigl(\lim_{y\to x(x_{0}>y_{0})} +\lim_{y \to x(x_{0}<y_{0})}\Bigr){\rm tr}(\gamma_{\mu}S^{(0)}_{F}(x,y)) \nonumber \\
  &=& -\frac{e}{2} \biggl( \sum_{n<-m} \bar{\psi}^{(-)}_{n}(x)\gamma_{\mu}\psi^{(-)}_{n}(x) - \sum_{n>-m} \bar{\psi}^{(+)}_{n}(x)\gamma_{\mu}\psi^{(+)}_{n}(x) \biggr),
\end{eqnarray}
where $iS^{(0)}_{F}(x,y)=\langle 0|T\hat{\psi}(x)\hat{\bar{\psi}}(y)|0 \rangle$.
In this expression, in the extarnal field free case two sums cancel, but for the field present case two sums do not cancel completely but diverge (it must be mentioned that if there is no external field and the system is homogeneous like the electron gas, the normal-ordered current gives same result with above definition). Hereby this result should be interpreted as describing the vacuum polarization effect. To obtain a finite value, we must perform the renormalization procedure. Now the HF current with the vacuum polarization becomes
\begin{eqnarray}
j^{HF}_{\mu}(x) &=& i\frac{e}{2}\Bigl(\lim_{y\to x(x_{0}>y_{0})}+\lim_{y\to x(x_{0}<y_{0})}\Bigr){\rm tr}(\gamma_{\mu}S^{HF}_{F}(x,y)) \nonumber \\
 &=& -e\sum_{-m<n\le F}\bar{\psi}^{(+)}_{n}(x)\gamma_{\mu}\psi^{(+)}_{n}(x) \nonumber \\
 & & -\frac{e}{2}\Biggl( \sum_{n<-m}\bar{\psi}^{(-)}_{n}(x)\gamma_{\mu}\psi^{(-)}_{n}(x) - \sum_{n>-m}\bar{\psi}^{(+)}_{n}(x)\gamma_{\mu}\psi^{(+)}_{n}(x)\Biggl).  
\end{eqnarray}

With these observations, we conclude that the HF total energy should be written by the HF currents, given by (19). For this purpose we must treat the convergence factor, which reflects time-ordering of field operators in the propagator, to be symmetric and use both the upper and lower halves of the integration contour. Using the following equation
\begin{eqnarray}
E &=& -i\int d^{3}x \int \frac{d\epsilon}{2\pi}\frac{1}{2}\Bigl(e^{i\epsilon\tau}\lim_{{\bf z}\to{\bf x}} + e^{-i\epsilon\tau}\lim_{{\bf z}\to{\bf x}}\Bigr)(-i\gamma\cdot\nabla+m_{0}+\gamma_{0}U(r)){\rm tr}(S^{HF}_{F}({\bf x},{\bf z},\epsilon)) \nonumber \\
 & & -\frac{1}{2}e^{2}\int d^{3}x \int d^{3}y \int \frac{d\epsilon_{1}}{2\pi} \int \frac{d\epsilon_{2}}{2\pi}D^{(0)}_{F}({\bf x}-{\bf y},0)_{\mu\nu} \nonumber \\
  & & \times \frac{1}{2}\Bigl(e^{i\epsilon_{1}\tau}\lim_{{\bf z}\to{\bf x}} +
e^{-i\epsilon_{1}\tau}\lim_{{\bf z}\to{\bf x}}\Bigr){\rm tr}(\gamma^{\mu}S^{HF}_{F}({\bf x},{\bf z},\epsilon_{1})) \nonumber \\
  & & \times \frac{1}{2}\Bigl(e^{i\epsilon_{2}\tau}\lim_{{\bf z'}\to{\bf y}} + e^{-i\epsilon_{2}\tau}\lim_{{\bf z'}\to{\bf y}}\Bigr){\rm tr}(\gamma^{\nu}S^{HF}_{F}({\bf y},{\bf z'},\epsilon_{2})) \nonumber \\
  & & +\frac{1}{2}e^{2}\int d^{3}x \int d^{3}y \int \frac{d\epsilon_{1}}{2\pi}\int\frac{d\epsilon_{2}}{2\pi}D^{(0)}_{F}({\bf x}-{\bf y},\epsilon_{1}-\epsilon_{2})_{\mu\nu} \nonumber \\
  & & \times {\rm tr}(\gamma^{\mu}S^{HF}_{F}({\bf x},{\bf y},\epsilon_{1})\frac{1}{2}(e^{i\epsilon_{1}\tau}+e^{-i\epsilon_{1}\tau})\gamma^{\nu}S^{HF}_{F}({\bf y},{\bf x},\epsilon_{2})\frac{1}{2}(e^{i\epsilon_{2}\tau'}+e^{-i\epsilon_{2}\tau'})), 
\end{eqnarray}
and performing the energy integration, we obtain ( in the Feynman gauge )
\begin{eqnarray}
E &=& \frac{1}{2}\int d^{3}x \lim_{{\bf z}\to{\bf x}}(-i\gamma\cdot\nabla+m_{0}+\gamma_{0}U(r)) \nonumber \\
  & & \times {\rm tr}\Bigl(\sum_{-\epsilon^{(-)}_{n}<-m}\psi^{(-)}_{n}({\bf x})\bar{\psi}^{(-)}_{n}({\bf z}) + \sum_{-m<\epsilon^{(+)}_{n}\le\epsilon_{F}}\psi^{(+)}_{n}({\bf x})\bar{\psi}^{(+)}_{n}({\bf z})-\sum_{\epsilon^{(+)}_{n}>\epsilon_{F}}\psi^{(+)}_{n}({\bf x})\bar{\psi}^{(+)}_{n}({\bf z}) \Bigr) \nonumber \\
  & & +\frac{1}{2}e^{2}\int d^{3}x \int d^{3}y \frac{g_{\mu\nu}}{4\pi|{\bf x}-{\bf y}|} \nonumber \\
  & & \times \frac{1}{2}\Biggl\{ {\rm tr} \gamma^{\mu}\Bigl(\sum_{-\epsilon^{(-)}_{n}<-m}\psi^{(-)}_{n}({\bf x})\bar{\psi}^{(-)}_{n}({\bf x}) + \sum_{-m<\epsilon^{(+)}_{n}\le\epsilon_{F}}\psi^{(+)}_{n}({\bf x})\bar{\psi}^{(+)}_{n}({\bf x})-\sum_{\epsilon^{(+)}_{n}>\epsilon_{F}}\psi^{(+)}_{n}({\bf x})\bar{\psi}^{(+)}_{n}({\bf x})\Bigr) \Biggr\} \nonumber \\
  & & \times \frac{1}{2}\Biggl\{ {\rm tr} \gamma^{\nu}\Bigl(\sum_{-\epsilon^{(-)}_{l}<-m}\psi^{(-)}_{l}({\bf y})\bar{\psi}^{(-)}_{l}({\bf y}) + \sum_{-m<\epsilon^{(+)}_{l}\le\epsilon_{F}}\psi^{(+)}_{l}({\bf y})\bar{\psi}^{(+)}_{l}({\bf y})-\sum_{\epsilon^{(+)}_{l}>\epsilon_{F}}\psi^{(+)}_{l}({\bf y})\bar{\psi}^{(+)}_{l}({\bf y})\Bigr) \Biggr\}  \nonumber \\
  & & -\frac{1}{2}e^{2}\int d^{3}x\int d^{3}y g_{\mu\nu}\frac{\exp(i|\epsilon^{(\pm)}_{n}-\epsilon^{(\pm)}_{l}||{\bf x}-{\bf y}|)}{4\pi|{\bf x}-{\bf y}|} \frac{1}{4}\Biggl\{ {\rm tr}\biggl( \nonumber \\
  & & \times \gamma^{\mu} \Bigl( \sum_{-\epsilon^{(-)}_{n}<-m}\psi^{(-)}_{n}({\bf x})\bar{\psi}^{(-)}_{n}({\bf y}) + \sum_{-m<\epsilon^{(+)}_{n}\le\epsilon_{F}}\psi^{(+)}_{n}({\bf x})\bar{\psi}^{(+)}_{n}({\bf y}) - \sum_{\epsilon^{(+)}_{n}>\epsilon_{F}}\psi^{(+)}_{n}({\bf x})\bar{\psi}^{(+)}_{n}({\bf y}) \Bigr) \nonumber \\
  & & \times \gamma^{\nu} \Bigl( \sum_{-\epsilon^{(-)}_{l}<-m}\psi^{(-)}_{l}({\bf y})\bar{\psi}^{(-)}_{l}({\bf x}) + \sum_{-m<\epsilon^{(+)}_{l}\le\epsilon_{F}}\psi^{(+)}_{l}({\bf y})\bar{\psi}^{(+)}_{l}({\bf x}) -\sum_{\epsilon^{(+)}_{l}>\epsilon_{F}}\psi^{(+)}_{l}({\bf y})\bar{\psi}^{(+)}_{l}({\bf x}) \Bigr)\biggr) \Biggr\}. \nonumber \\
  & &  
\end{eqnarray}
In (21), we handle not only $N$-electrons but infinite number of 
interacting electrons and positrons. It is essentially different from the non-relativistic theories. The direct term and the exchange term give the quantum
correction self-consistently. In the direct term, there is no energy-flow and the photon propagator becomes an instantaneous Coulomb interaction, while for the exchange term the exponent in the propagator has the energy difference and it describes the retardation effect. Both the Dirac sea and the Fermi sea contain
the Hartree-Fock quasi-particles. 
   
The (21) involves two types of divergences. One is the zero-point energy of
the fermion system, while other has its origin in the interaction between particles.
In principle, the zero-point energy can be removed by a simple subtraction, 
$E_{HF} = E - E_{ZP}$, and it gives the redefinition of the HF total energy. 
Here we use the next definition for the zero-point energy
\begin{eqnarray}
E_{ZP} &=& -i\int d^{3}x\int\frac{d\epsilon}{2\pi}\frac{1}{2}(e^{i\epsilon\tau} \lim_{{\bf z}\to{\bf x}}+e^{-i\epsilon\tau}\lim_{{\bf z}\to{\bf x}})(-i\gamma\cdot\nabla+m_{0}){\rm tr}S^{(0)}_{F}({\bf x},{\bf z},\epsilon),
\end{eqnarray}
where $iS^{(0)}_{F}(x,y)=\langle 0|T\hat{\psi}(x)\hat{\bar{\psi}}(y)|0\rangle$ is the free, and no-external-field fermion propagator. 
On the other hand, the divergences which originate from the interaction must be
treated by the renormalization methods in the quantum field theory\cite{4}. For this
purpose we should introduce some reasonable cut-off, and prepare the 
renormalization constants. But in the above case we can not apply the commonly used
method for the field theory. There is a large number of efforts to perform the renormalization in atomic QED calculations\cite{3}. All of them are based on the perturbative QED. We have not met any of non-perturbative renormalization in QED, which can be applied to our thoery. This problem still remains as an open question.

\section{Operator Formalism}
\label{sec:ope}

The most part of QED in literature treats the translation and Lorentz invariant vacuum, and which contains no external, central field. It is constructed by 
the field operators and Green's functions. 
On the other hand, the many-body theory is 
constructed not only by the field operators and Green's functions, 
but also by the creation-annihilation operators, 
especially in the atomic and moleculer physics. 
In this formalism, we frequently use the occupation numbers for the 
one-particle states or the density matrices, 
and also the canonical transformation among the creation-annihilation operators
take many active parts. 
These operators and the Hamiltonian are in the Schr\"{o}dinger representation. 
We can find a large number of approximation schemes in this formalism, and they have various advantageous features. 
To combine these methods to QED, we investigate the creation-annihilation operator formalism for QED in this section.

\subsection{Hamiltonian}
This subsection is devoted to introduce a fully normal-ordered relativistic Hamiltonian for our theory. We essentially need a normal-ordered Hamiltonian, to use operator methods developed in the nonrelativistic many-body theory. We start from (8). This Hamiltonian is written in the Heisenberg 
representation, containing no approximation. The field operators are now substituted into the expanded form given in (11) and (12). 
For the kinetic energy term we can simply 
take the normal-ordering, and neglecting a c-number. Also the time is chosen as $x_{0}=0$ to become the Schr\"{o}dinger representation.
About the interaction term, the photon propagator is substituted by the free propagator
and we perform the Fourier transformation in time $y_{0}$, and then we transform to the Schr\"{o}dinger representation at the time $x_{0}=0$. We use the normal-ordered currents and take a product of them to express the interaction term. Once more we employ normal-ordering for this product, and we obtain 16 fully 
normal-ordered operators with 8 one-body operators. We simply 
negrect the latter.
Then we obtain the next fully normal-ordered relativistic Hamiltonian
\begin{eqnarray}
\hat{H} &=& \hat{H}_{K} + \hat{H}_{I},  \\
\hat{H}_{K} &=& \sum_{i,j} T^{\pm\pm}_{ij}( \hat{a}^{\dagger}_{i}\hat{a}_{j} + \hat{a}^{\dagger}_{i}\hat{b}^{\dagger}_{j} + \hat{b}_{i}\hat{a}_{j} - \hat{b}^{\dagger}_{j}\hat{b}_{i} ), \\
\hat{H}_{I} &=& \frac{1}{2} \sum_{i,j,k,l} V^{\pm\pm\pm\pm}_{ijkl}(\hat{a}^{\dagger}_{i}\hat{a}^{\dagger}_{k}\hat{a}_{l}\hat{a}_{j} + \hat{a}^{\dagger}_{i}\hat{a}^{\dagger}_{k}\hat{b}^{\dagger}_{l}\hat{a}_{j} + \hat{a}^{\dagger}_{i}\hat{a}_{j}\hat{b}_{k}\hat{a}_{l} + \hat{a}^{\dagger}_{i}\hat{b}^{\dagger}_{l}\hat{a}_{j}\hat{b}_{k} \nonumber \\
 & & + \hat{a}^{\dagger}_{i}\hat{b}^{\dagger}_{j}\hat{a}^{\dagger}_{k}\hat{a}_{l} + \hat{a}^{\dagger}_{i}\hat{b}^{\dagger}_{j}\hat{a}^{\dagger}_{k}\hat{b}^{\dagger}_{l} + \hat{a}^{\dagger}_{i}\hat{b}^{\dagger}_{j}\hat{b}_{k}\hat{a}_{l} + \hat{a}^{\dagger}_{i}\hat{b}^{\dagger}_{l}\hat{b}^{\dagger}_{j}\hat{b}_{k} + \hat{a}^{\dagger}_{k}\hat{b}_{i}\hat{a}_{j}\hat{a}_{l} + \hat{a}^{\dagger}_{k}\hat{b}^{\dagger}_{l}\hat{b}_{i}\hat{a}_{j} \nonumber \\
 & & + \hat{b}_{i}\hat{a}_{j}\hat{b}_{k}\hat{a}_{l} + \hat{b}^{\dagger}_{l}\hat{a}_{j}\hat{b}_{i}\hat{b}_{k} + \hat{b}^{\dagger}_{j}\hat{a}^{\dagger}_{k}\hat{b}_{i}\hat{a}_{l} + \hat{a}^{\dagger}_{k}\hat{b}^{\dagger}_{j}\hat{b}^{\dagger}_{l}\hat{b}_{i} + \hat{b}^{\dagger}_{j}\hat{b}_{k}\hat{b}_{i}\hat{a}_{l} + \hat{b}^{\dagger}_{j}\hat{b}^{\dagger}_{l}\hat{b}_{k}\hat{b}_{i}), 
\end{eqnarray}
where we designate matrix elements as
\begin{eqnarray}
T^{\pm\pm}_{ij} &=& \int d^{3}x \bar{\psi}^{(\pm)}_{i}({\bf x})(-i\gamma\cdot\nabla + m_{0} + \gamma_{0}U(r))\psi^{(\pm)}_{j}({\bf x}),
\end{eqnarray}
\begin{eqnarray}
V^{\pm\pm\pm\pm}_{ijkl} &=& \frac{1}{2}e^{2}\int d^{3}x \int d^{3}y \bar{\psi}^{(\pm)}_{i}({\bf x})\gamma^{\mu}\psi^{(\pm)}_{j}({\bf x})\frac{g_{\mu\nu}\exp(i\Delta\epsilon|{\bf x}-{\bf y}|)}{4\pi|{\bf x}-{\bf y}|} \bar{\psi}^{(\pm)}_{k}({\bf y})\gamma^{\nu}\psi^{(\pm)}_{l}({\bf y}).  \nonumber \\
& & 
\end{eqnarray}
Here, $i,j,k,l$ denote quantum numbers for one-particle states, and $\Delta\epsilon\equiv|\epsilon_{k}-\epsilon_{l}| = |\epsilon_{i}-\epsilon_{j}|$. Hereafter we use this Hamiltonian in our theory. It must be noted that this Hamiltonian (Eq.(23)) is in the Schr\"{o}dinger representation.
This Hamiltonian contains operators that they  do not conserve the particle number, but conserve the total charge of the system, as the usual relativistic case. 

Next, we try to evaluate the expectation value of (23) for the Fermi sea, which is defined as
\begin{eqnarray}
|F\rangle &=& \prod_{-m<n\le F} \hat{a}^{\dagger}_{n}|0\rangle. 
\end{eqnarray}
It is clear that the naive application of usual nonrelativistic method, such as performing the factrization and taking the pairing, does not give the expression given in (21). Then here we propose to examine the relation between the contraction and the pairing. The contraction and pairing are defined by
\begin{eqnarray}
A^{\bullet}(t)B^{\bullet}(t') &=& T(A(t)B(t')) -  N(A(t)B(t')), \\
A^{*}(t)B^{*}(t') &=& A(t)B(t') - N(A(t)B(t')),
\end{eqnarray}
here $\bullet$ denotes a contraction, $*$ denotes a pairing. The normal-order $N$ is refered to the Fermi sea. The relation between the contraction and the pairings is
\begin{eqnarray}
A^{\bullet}(t)B^{\bullet}(t') &=& A^{*}(t)B^{*}(t')\theta(t-t')-B^{*}(t')A^{*}(t)\theta(t'-t). 
\end{eqnarray}
The definition of the pairing gives $\langle F|A(t)B(t')|F \rangle = A^{*}(t)B^{*}(t')$, the contraction becomes
\begin{eqnarray}
A^{\bullet}(t)B^{\bullet}(t') &=& \langle F|A(t)B(t')|F \rangle \theta(t-t') - \langle F|B(t')A(t)|F \rangle \theta(t'-t). 
\end{eqnarray}
Especially in the case of $t=t'$
\begin{eqnarray}
\langle F|T(A(t)B(t'))|F \rangle _{t=t'} &=& \langle F|A^{\bullet}(t)B^{\bullet}(t')+N(A(t)B(t'))|F \rangle _{t=t'} \nonumber \\
&=& \langle F|A(t)B(t')|F \rangle \theta(t-t')|_{t=t'} - \langle F|B(t')A(t)|F \rangle \theta(t'-t)|_{t=t'}. \nonumber \\
& & 
\end{eqnarray}
Now we introduce the definition for the step function at the same times as
\begin{eqnarray}
\theta(t-t')_{t=t'} &=& \theta(t'-t)_{t=t'} = \frac{1}{2}. 
\end{eqnarray}
By this definition we obtain the following relation:
\begin{eqnarray}
\langle F|T(A(t)B(t'))|F \rangle _{t=t'} &=& \langle F|A^{\bullet}(t)B^{\bullet}(t')|F \rangle _{t=t'} \nonumber \\
 &=& \frac{1}{2}\langle F|A(t)B(t)|F \rangle -\frac{1}{2}\langle F|B(t)A(t)|F \rangle. 
\end{eqnarray}
With the examination given above, we introduce a hypothesis that the Hamiltonian given in (23) is written by operators in the same-time $T$-products, and when we factorize the vacuum expectation value for an operator product with the aid of the Wick theorem,
each contraction should be calculated by the definition given in (35).
 
By using this assumption, for example
\begin{eqnarray}
\sum_{ij}\hat{a}^{\dagger\bullet}_{i}\hat{a}^{\bullet}_{j} &=& \sum_{ij}\frac{1}{2} \Bigl( \langle F|\hat{a}^{\dagger}_{i}\hat{a}_{j}|F \rangle - \langle F|\hat{a}_{j}\hat{a}^{\dagger}_{i}|F \rangle \Bigr) = \frac{1}{2}(\sum_{-m<i \le F} - \sum_{i>F}) \delta_{ij}, \\
\sum_{ij}-\hat{b}^{\dagger\bullet}_{j}\hat{b}^{\bullet}_{i} &=& \sum_{ij}-\frac{1}{2}\Bigl( \langle F|\hat{b}^{\dagger}_{j}\hat{b}_{i}|F \rangle - \langle F|\hat{b}_{i}\hat{b}^{\dagger}_{j}|F \rangle \Bigr) = \frac{1}{2}\sum_{i<-m} \delta_{ij}, \\
\sum_{ijkl}\hat{a}^{\dagger\bullet}_{i}\hat{a}^{\dagger\bullet\bullet}_{k}\hat{a}^{\bullet}_{l}\hat{a}^{\bullet\bullet}_{j} &=& \sum_{ijkl}-\hat{a}^{\dagger\bullet}_{i}\hat{a}^{\bullet}_{l}\hat{a}^{\dagger\bullet\bullet}_{k}\hat{a}^{\bullet\bullet}_{i} \nonumber \\
 &=& \sum_{ijkl}-\frac{1}{2}\Bigl( \{ \langle F|\hat{a}^{\dagger}_{i}\hat{a}_{l}|F \rangle - \langle F|\hat{a}_{l}\hat{a}^{\dagger}_{i}|F \rangle \}\Bigr)\frac{1}{2}\Bigl( \{ \langle F|\hat{a}^{\dagger}_{k}\hat{a}_{j}|F \rangle - \langle F|\hat{a}_{j}\hat{a}^{\dagger}_{k}|F \rangle \}\Bigr) \nonumber \\
 &=&  -\frac{1}{4}\{\sum_{-m<i\le F} - \sum_{i>F}\}\delta_{il}\{\sum_{-m<k\le F}-\sum_{k>F} \}\delta_{kj}, 
\end{eqnarray}
and other terms are also calculated. The final result becomes
\begin{eqnarray}
\langle F|\hat{H}_{K}|F \rangle &=& \frac{1}{2}\sum_{i<-m} T^{--}_{ii} + \frac{1}{2}(\sum_{-m<i \le F} - \sum_{i>F} )T^{++}_{ii}, \\
\langle F|\hat{H}_{I}|F \rangle &=& \frac{1}{8}\biggl\{ (\sum_{-m<i\le F}-\sum_{i>F})(\sum_{-m<k \le F}-\sum_{k>F})(V^{++++}_{iikk}-V^{++++}_{ikki}) \nonumber \\
 & & +(\sum_{-m<i \le F}-\sum_{i>F})(\sum_{k<-m})(V^{++--}_{iikk}-V^{+--+}_{ikki}) \nonumber \\
 & & +(\sum_{-m<k\le F}-\sum_{k>F})(\sum_{i<-m})(V^{--++}_{iikk}-V^{-++-}_{ikki}) \nonumber \\
 & & +(\sum_{i<-m})(\sum_{k<-m})(V^{----}_{iikk}-V^{----}_{ikki}) \biggr\}.
\end{eqnarray}
Now we can introduce the HF equations in the following form:
\begin{eqnarray}
T^{++}_{ii} + \frac{1}{2}(\sum_{k<-m}+\sum_{-m<k \le F}-\sum_{k>F})(V^{++\pm\pm}_{iikk}-V^{+\pm\pm+}_{ikki}) &=& +\epsilon^{(+)}_{i}, \\
T^{--}_{ii} + \frac{1}{2}(\sum_{k<-m}+\sum_{-m<k \le F}-\sum_{k>F})(V^{--\pm\pm}_{iikk}-V^{-\pm\pm-}_{ikki}) &=& -\epsilon^{(-)}_{i}.
\end{eqnarray}
These equations are rederived later, where we discuss the HF condition, and we 
can confirm that the definition of the HF equation given above has no conflict with it. The HF Hamiltonian can easily be determined by use of the above 
results. By using (41) and (42), we obtain the HF total energy
\begin{eqnarray}
E_{HF} &=& \frac{1}{2}\Bigl\{ \sum_{i<-m}(-\epsilon^{(-)}_{i}) + \sum_{-m<i\le F}\epsilon^{(+)}_{i}  - \sum_{i>F}\epsilon^{(+)}_{i} \Bigr\} \nonumber \\
 & & -\frac{1}{8}(\sum_{i<-m}+\sum_{-m<i\le F}-\sum_{i>F})(\sum_{j<-m}+\sum_{-m<j\le F}-\sum_{j>F})(V^{\pm\pm\pm\pm}_{iijj}-V^{\pm\pm\pm\pm}_{ijji}) \nonumber \\
 & & - E_{ZP}. 
\end{eqnarray}
Here we give the zero-point energy as in the (22), and it can also be calculated by the same way in this subsection.

Eq.(43) gives a consistent result with that derived by the Green's function method in the last section. The results given above imply that we can generalize this method to calculate expectation values for arbitrary products of operators. We always factorize the matrix elements of an operator product by using the Wick theorem into the sum of the products of contractions, and the contraction is calculated after the definition given above. It is clear from the above discussion that {\it only} the definition of the contraction 
for same-time operators is modified. Later we will use this definition to derive the HF, TDHF and RPA. It must be taken into account the fact 
that the operator method based on the Schr\"{o}dinger representation, i.e.,
all operators appearing in the theory obey the same-time, $t=0$.

Now we assert that we obtain a basis to construct the relativistic many-body theory in the operator formalism. The Schr\"{o}dinger equation in our theory becomes
\begin{eqnarray}
|\Psi(t)\rangle &=& \sum_{N}\sum_{I}A^{N}_{I}(t)\Phi^{N}_{I}({\bf x}_{1},{\bf x}_{2},...,{\bf x}_{N}), \\
 & & i\frac{\partial}{\partial t}|\Psi(t)\rangle = \hat{H}|\Psi(t)\rangle, 
\end{eqnarray}
where $\Phi^{N}_{I}$ is an $N$-particle Slater determinant, and $A^{N}_{I}(t)$ is a time-dependent coefficient. It is clear from the above derivation, the Hamiltonian and the wavefunction depend on the gauge. A problem, how we should be treat the gauge dependence of our theory in many-body procedure, still remains.

\subsection{The Slater determinant}

In the Hartree-Fock theory, the many-particle wavefunction is expressed by a single Slater determinant, and the one-particle functions are optimized variationally. It is well-known that any Slater determinant can be written in the Thouless form\cite{13}
\begin{eqnarray}
|\Phi\rangle &=& e^{i\hat{G}(z)}|\Phi_{0}\rangle, \\
\hat{G}(z) &=& \sum_{ij}z_{ij}\hat{c}^{\dagger}_{i}\hat{c}_{j},
\end{eqnarray}
where $|\Phi_{0}\rangle$ is a fixed $N$-particle determinantal state. The parameter $z_{ij}$ have to satisfies the Hermitian condition, $z_{ij}=z^{*}_{ji}$, as $e^{i\hat{S}}$ will be an unitary operator. The exponent generates a canonical transformation on fermion operators. This second quantized representation of a single Slater determinant is very useful in operator methods. We try to generalize it to a relativistic form, because it plays essential roles in the later discussions. 

Let us introduce several Dirac vacua
\begin{eqnarray}
\hat{a}_{n}|0\rangle &=& \hat{b}_{n}|0\rangle = 0, \\
\hat{a}'_{n}|0'\rangle &=& \hat{b}'_{n}|0'\rangle = 0, \\
\hat{a}^{HF}_{n}|0_{HF}\rangle &=& \hat{b}^{HF}_{n}|0_{HF}\rangle = 0,  
\end{eqnarray}
where the first, second and third is old, new and HF vacuum, respectively.
Determinants are given as
\begin{eqnarray}
|\Phi_{0}\rangle &=& \prod_{n} \hat{a}^{\dagger}_{n}|0\rangle, \\
|\Phi'_{0}\rangle &=& \prod_{n} \hat{a}'^{\dagger}_{n}|0'\rangle. 
\end{eqnarray}
We want to obtain a relativistic exponential operator which transforms one to another representations.
We assume the following relations for the canonical transformation:
\begin{eqnarray}
\hat{a}'_{m} &=& e^{i\hat{S}}\hat{a}_{m}e^{-i\hat{S}}  =  \sum_{n}(\alpha_{mn}\hat{a}_{n} + \beta_{mn}\hat{b}^{\dagger}_{n}), \\
\hat{a}'^{\dagger}_{m} &=& e^{i\hat{S}}\hat{a}^{\dagger}_{m}e^{-i\hat{S}} = \sum_{n}(\alpha^{*}_{mn}\hat{a}^{\dagger}_{n} + \beta^{*}_{mn}\hat{b}_{n}), \\
\hat{b}'^{\dagger}_{m} &=& e^{i\hat{S}}\hat{b}^{\dagger}_{m}e^{-i\hat{S}} = \sum_{n}(\beta_{mn}\hat{a}_{n} + \gamma_{mn}\hat{b}^{\dagger}_{n}), \\
\hat{b}'_{m} &=& e^{i\hat{S}}\hat{b}_{m}e^{-i\hat{S}} = \sum_{n}(\beta^{*}_{mn}\hat{a}^{\dagger}_{n} + \gamma^{*}_{mn}\hat{b}_{n}), 
\end{eqnarray}
where $\hat{S}^{\dagger}=\hat{S}$ will be satisfied. Here new operators are expanded by a complete set of the old representation. The definition of the new vacuum becomes
\begin{eqnarray}
\hat{a}'_{n}|0'\rangle &=& (e^{i\hat{S}}\hat{a}_{n}e^{-i\hat{S}})e^{i\hat{S}}|0\rangle = 0. 
\end{eqnarray}
Therefore, the new vacuum is formally given by a superposition of the states in the old representaion. In the relativistic theory, the vacuum also be transformed. The relation of two determinants becomes 
\begin{eqnarray}
|\Phi'_{0}\rangle = \prod_{n}\hat{a}'^{\dagger}_{n}|0'\rangle = \prod_{n}(e^{i\hat{S}}\hat{a}^{\dagger}_{n}e^{-i\hat{S}})e^{i\hat{S}}|0\rangle = e^{i\hat{S}}\prod_{n}\hat{a}^{\dagger}_{n}|0\rangle = e^{i\hat{S}}|\Phi_{0}\rangle. 
\end{eqnarray}
Now we introduce the exponential operator in the following form
\begin{eqnarray}
e^{i\hat{S}} &=& \exp\Bigl\{i\sum_{mn}(\alpha_{mn}\hat{a}^{\dagger}_{m}\hat{a}_{n} + \beta_{mn}\hat{a}^{\dagger}_{m}\hat{b}^{\dagger}_{n} + \beta^{*}_{mn}\hat{b}_{n}\hat{a}_{m} + \gamma^{*}_{mn}\hat{b}_{n}\hat{b}^{\dagger}_{m})\Bigr\}.
\end{eqnarray}
Here we demand that the matrices formed by the paramerters $\alpha_{mn},\beta_{mn},\beta^{*}_{mn},\gamma^{*}_{mn}$ should be Hermitian. 
Then the exponential operator formally given above is unitary. 
It is clear that the (59) satisfies relations 
given by (53)$\sim$(56). Each parameter should be determined 
in the approximation scheme, and in the HF theory they are variationally determined and reflect the variational condition. 
Variations for one-particle functions are generated by each operator product 
in the exponent. It must be noticed that the second and the third terms in the exponent can be understood as the generalized Bogoliubov transformation to the Dirac vacuum, which does not conserve the particle numbers, and they describe the fluctuations of the Dirac vacuum. In the HF theory, the parameters included in them would be satisfied not only the HF variational condition but also vanish the matrix elements for the parts of the Hamiltonian which do not conserve particle numbers. These conditions must be satisfied for the diagonalization of the Hamiltonian. In (53)$\sim$(56), the new operators can be interpreted 
as quasi-particles, and they are given by superpositions of old operators. 
Later we use this exponential operator to derive the TDHF theory, the HF condition and the RPA equations. The most important point in this exponential operator is that it gives the degrees of freedom of a system explicitly. By this feature we can introduce many flexible approximation scheme, and can describe the dynamics of a system in detail, much more than the usual QED treatment.

Now we must mention some mathematical points characteristic in the field theory. It is well-known fact that, in the theory of infinite degrees of freedom as quantum field theory, a different Fock space construct a separated Hilbert space, and two representations are unitarily inequivalent to each other\cite{14}. 
Thus expansions used in (57) and (58) give just formal definitions. A vector of one representation can not be expanded by bases of another representation. For example, a state for the observed electrons can not be expanded meaningfully by the bases of the bare electron Hilbert space. Thus we can not treat the 
interaction effect by only one Hilbert space. 
In fact, the renormalization procedure is a transformation to another representation. The renormalization method projects bare parameters to physical one, and transforms to the Fock space of observed particles. Therefore we interpret that a state of a new representation (the HF state) is {\it formally} expanded by old bases by using $e^{i\hat{S}}$, then we evaluate matrix elements, and have to employ the renormalization to transform to an observed particle (Hartree-Fock quasi-particle) space. In our theory, the Fock space for HF states is self-consistently selected with the renormalization procedure.  
 
Using the exponential operator given in (59), we can write the expectation value of the Hamiltonian (Eq.(23)) in a general Slater determinant (Eq.(58)) 
into an expanded form, by the same way as nonrelativistic cases. 
Here we will write a relativistic Slater determinat as
\begin{eqnarray}
|\Phi(\alpha,\beta,\beta^{*},\gamma^{*})\rangle = e^{i\hat{S}(\alpha,\beta,\beta^{*},\gamma^{*})}|\Phi_{0}\rangle .
\end{eqnarray}
The expectation value of our Hamiltonian is given as follows:
\begin{eqnarray}
\langle \Phi(\alpha,\beta,\beta^{*},\gamma^{*})|\hat{H}|\Phi(\alpha,\beta,\beta^{*},\gamma^{*})\rangle &=& \langle \Phi_{0}|\hat{H}|\Phi_{0} \rangle \nonumber \\ 
 & & + i\langle \Phi_{0}|[\hat{H},\hat{S}(\alpha,\beta,\beta^{*},\gamma^{*})]|\Phi_{0}\rangle  \nonumber \\
 & & + \frac{i^{2}}{2!}\langle\Phi_{0}|[[\hat{H},\hat{S}(\alpha,\beta,\beta^{*},\gamma^{*})],\hat{S}(\alpha,\beta,\beta^{*},\gamma^{*})]|\Phi_{0}\rangle \nonumber \\
 & &  + {\cal O}(\hat{S}^{3}). 
\end{eqnarray}
In (61), the first line is the HF total energy, the second line gives the first derivatives with respect to the parameters ($\alpha,\beta,\beta^{*},\gamma^{*}$), and it must be zero in the HF condition. The third line corresponds
to the second derivatives and they determine the stability of the HF state, it 
is equivalent to the stability of collective modes in the RPA. It is obvious 
fact that the operator formalism makes possible to do these discussions. By using our theory, for example, it may be possible to discuss the instability of the HF solution, or to apply the coupled-cluster method in the electronic structure theory.

\subsection{The time-dependent Hartree-Fock theory}

In this and in the following subsection, we will derive the TDHF, HF condition and RPA, by using the operator method, based on the previous discussions. Here it is also the aim to demonstrate the usefulness of our theory. We take the starting point to the TDHF, the theory for the time evolution of the mean-field\cite{15}.

Using the Hamiltonian ( Eq.(23)), we construct the action functional
\begin{eqnarray}
S_{HF} &=& \int^{t_{f}}_{t_{i}}dt \langle \Phi(\alpha(t),\beta(t),\beta^{*}(t),\gamma^{*}(t))|i\frac{\partial}{\partial t}-\hat{H}|\Phi(\alpha(t),\beta(t),\beta^{*}(t),\gamma^{*}(t))\rangle
\end{eqnarray}
Here the time-dependence of the determinantal state is borne on the parameter set ($\alpha,\beta,\beta^{*},\gamma^{*}$). Introduce the Dirac-Frenkel variation principle,
\begin{eqnarray}
\delta S_{HF} &=& \delta\int^{t_{f}}_{t_{i}} dt \langle \Phi(\alpha(t),\beta(t),\beta^{*}(t),\gamma^{*}(t))|i\frac{\partial}{\partial t}-\hat{H}|\Phi(\alpha(t),\beta(t),\beta^{*}(t),\gamma^{*}(t))\rangle = 0. \nonumber \\
 & & 
\end{eqnarray}
Thus we obtain the stationary condition
\begin{eqnarray}
0 &=& \delta \langle\Phi(\alpha,\beta,\beta^{*},\gamma^{*})|i\frac{\partial i\hat{S}}{\partial t}-\hat{H}|\Phi(\alpha,\beta,\beta^{*},\gamma^{*})\rangle.   
\end{eqnarray}
Therefore we take the derivatives with respect to the parameters and derive a set of TDHF equations
\begin{eqnarray}
0 &=& \langle \Phi(\alpha,\beta,\beta^{*},\gamma^{*})|[\hat{a}^{\dagger}_{m}\hat{a}_{n},i\frac{\partial i\hat{S}}{\partial t}-\hat{H}]|\Phi(\alpha,\beta,\beta^{*},\gamma^{*})\rangle, \\
0 &=& \langle \Phi(\alpha,\beta,\beta^{*},\gamma^{*})|[\hat{a}^{\dagger}_{m}\hat{b}^{\dagger}_{n},i\frac{\partial i\hat{S}}{\partial t}-\hat{H}]|\Phi(\alpha,\beta,\beta^{*},\gamma^{*})\rangle, \\
0 &=& \langle \Phi(\alpha,\beta,\beta^{*},\gamma^{*})|[\hat{b}_{m}\hat{a}_{n}, i\frac{\partial i\hat{S}}{\partial t}-\hat{H}]|\Phi(\alpha,\beta,\beta^{*},\gamma^{*})\rangle, \\
0 &=& \langle \Phi(\alpha,\beta,\beta^{*},\gamma^{*})|[\hat{b}_{m}\hat{b}^{\dagger}_{n},i\frac{\partial i\hat{S}}{\partial t}-\hat{H}]|\Phi(\alpha,\beta,\beta^{*},\gamma^{*})\rangle, 
\end{eqnarray}
where 
\begin{eqnarray}
i\frac{\partial i\hat{S}}{\partial t} = -\sum_{mn}(\dot{\alpha}_{mn}\hat{a}^{\dagger}_{m}\hat{a}_{n} + \dot{\beta}_{mn}\hat{a}^{\dagger}_{m}\hat{b}^{\dagger}_{n} + \dot{\beta}^{*}_{mn}\hat{b}_{n}\hat{a}_{m} + \dot{\gamma}^{*}_{mn}\hat{b}_{n}\hat{b}^{\dagger}_{m}).
\end{eqnarray}
Another form can be obtained for the TDHF equations. We also derive the TDHF equations in the density matrix form because later we use them for the derivation of the HF condition and RPA equations. Hereafter we write $|\Phi(\kappa)\rangle$ for $|\Phi(\alpha,\beta,\beta^{*},\gamma^{*})\rangle$. The definition of the density matrix is
\begin{equation}
\rho(\kappa) = \left(
        \begin{array}{cccc}
            \rho^{++}_{nm}(\kappa) & \rho^{+-}_{nm}(\kappa) \\
            \rho^{-+}_{nm}(\kappa) & \rho^{--}_{nm}(\kappa) 
        \end{array}
        \right), 
\end{equation}
where
\begin{eqnarray}
\rho^{++}_{nm}(\kappa) &=& \langle \Phi(\kappa)|\hat{a}^{\dagger}_{m}\hat{a}_{n}|\Phi(\kappa) \rangle  \nonumber \\
 &=& \langle \Phi_{0}|\hat{a}^{\dagger\bullet}_{m}\hat{a}^{\bullet}_{n}|\Phi_{0} \rangle + i\langle \Phi_{0}|[\hat{a}^{\dagger}_{m}\hat{a}_{n},\hat{S}]|\Phi_{0} \rangle + \frac{i^{2}}{2}\langle \Phi_{0}|[[\hat{a}^{\dagger}_{m}\hat{a}_{n},\hat{S}],\hat{S}]|\Phi_{0} + \cdots \nonumber \\
 &=& \rho^{++(0)}_{nm} + \rho^{++(1)}_{nm}(\kappa) + \rho^{++(2)}_{nm}(\kappa) + \cdots  ,
\end{eqnarray}
where an expansion at a specific point is given. The definition for others are, respectively, 
\begin{eqnarray}
\rho^{-+}_{nm}(\kappa) &=& \langle \Phi(\kappa)|\hat{a}^{\dagger}_{m}\hat{b}^{\dagger}_{n}|\Phi(\kappa) \rangle, \\
\rho^{+-}_{nm}(\kappa) &=& \langle \Phi(\kappa)|\hat{b}_{m}\hat{a}_{n}|\Phi(\kappa) \rangle, \\
\rho^{--}_{nm}(\kappa) &=& \langle \Phi(\kappa)|\hat{b}_{m}\hat{b}^{\dagger}_{n}|\Phi(\kappa) \rangle.   
\end{eqnarray}
It must be noted that $\hat{a}^{\dagger}_{n}\hat{a}_{m},\hat{a}^{\dagger}_{n}\hat{b}^{\dagger}_{m},\hat{b}_{n}\hat{a}_{m},\hat{b}_{n}\hat{b}^{\dagger}_{m}$ construct a Lie algebra, so it is not necessary to concern ourselves about other types of operator-pairs throughout our theory. Using the relations
\begin{eqnarray}
\frac{\partial}{\partial t}\rho^{++}_{nm}(\kappa) &=& \langle \Phi(\kappa)|[\hat{a}^{\dagger}_{m}\hat{a}_{n},\frac{\partial i\hat{S}}{\partial t}]|\Phi(\kappa) \rangle, \\
\frac{\partial}{\partial i\alpha_{mn}}\langle \Phi(\kappa)|\hat{H}|\Phi(\kappa) \rangle &=& \langle \Phi(\kappa)|[\hat{H},\hat{a}^{\dagger}_{m}\hat{a}_{n}]|\Phi(\kappa) \rangle, 
\end{eqnarray}
and so on, we obtain the TDHF equations in the following forms:
\begin{eqnarray}
i\frac{\partial}{\partial t}\rho^{++}_{nm}(\kappa) &=& -\frac{\partial}{\partial i\alpha_{mn}}\langle \Phi(\kappa)|\hat{H}|\Phi(\kappa) \rangle, \\
i\frac{\partial}{\partial t}\rho^{-+}_{nm}(\kappa) &=& -\frac{\partial}{\partial i\beta_{mn}}\langle \Phi(\kappa)|\hat{H}|\Phi(\kappa) \rangle, \\
i\frac{\partial}{\partial t}\rho^{+-}_{nm}(\kappa) &=& -\frac{\partial}{\partial i\beta^{*}_{nm}}\langle \Phi(\kappa)|\hat{H}|\Phi(\kappa) \rangle, \\
i\frac{\partial}{\partial t}\rho^{--}_{nm}(\kappa) &=& -\frac{\partial}{\partial i\gamma^{*}_{nm}}\langle \Phi(\kappa)|\hat{H}|\Phi(\kappa) \rangle. 
\end{eqnarray}

In our formalism, theories of beyond-TDHF also can easily be constructed.
For example, we can introduce a multi-configurational wavefunction $|\Psi(t)\rangle = \sum_{I}A_{I}(t)|\Phi_{I}\rangle $ (here $|\Phi_{I}\rangle$ is a Slater determinant) for making an action functional and adopt the Dirac-Frenkel variation. Then we obtain the multi-configurational TDHF in QED
\begin{eqnarray}
\delta S_{MC} &=& \delta  \int dt \sum_{IJ}A^{*}_{I}(t)A_{J}(t)\langle \Phi_{I}|e^{-i\hat{S}(\alpha,\beta,\beta^{*},\gamma^{*})}\Bigl(i\frac{\partial}{\partial t}-\hat{H}\Bigr)e^{i\hat{S}(\alpha,\beta,\beta^{*},\gamma^{*})}|\Phi_{J}\rangle,
\end{eqnarray} 
where the configuration coefficients and one-particle functions are optimized simultaneously.

\subsection{The HF condition and the random phase approximation}
The TDHF gives a picture that the presentation point for the system moves in a parameter space. The RPA has the meaning of the linearized approximation to the TDHF. It is well-known that the RPA equation can be derived by several different methods. The famous equation-of-motion method developed by Sawada\cite{16} can also be applied to our case, but in our case it is too cumbersome. Here we use the TDHF equations and derive the RPA equations by expanding the density matrices\cite{13}. This method is convenient to derive the RPA and also we can easily interpret the physics and relations to other theories.

First, we try to write down the expectation value of the Hamiltonian (Eq.(23)) 
in a single determinant $|\Phi_{0}\rangle$. It is given as follows
\begin{eqnarray}
\langle \Phi(\kappa)|\hat{H}|\Phi(\kappa) \rangle &=& \sum_{ij}T^{\pm\pm}_{ij}\{\rho^{++}_{ji}(\kappa)+\rho^{+-}_{ji}(\kappa)+\rho^{-+}_{ji}(\kappa)+\rho^{--}_{ji}(\kappa)\} \nonumber \\
 & & + \frac{1}{2}\sum_{ijkl}(V^{\pm\pm\pm\pm}_{ijkl}-V^{\pm\pm\pm\pm}_{ilkj}) \nonumber \\
 & & \times \{\rho^{++}_{ji}(\kappa)+\rho^{+-}_{ji}(\kappa)+\rho^{-+}_{ji}(\kappa)+\rho^{--}_{ji}(\kappa)\} \nonumber \\
 & & \times \{\rho^{++}_{lk}(\kappa)+\rho^{+-}_{lk}(\kappa)+\rho^{-+}_{lk}(\kappa)+\rho^{--}_{lk}(\kappa)\}.
\end{eqnarray}
Next, we introduce an exponential operator to use the expansion of the density matrix by parameters. This operator must explicitly divide the one-particle space into above and below the fermi level. Then we introduce the next form
\begin{eqnarray}
e^{i\hat{S}} &=& \exp\Biggl\{ i \biggl(\sum_{\mu\nu}\alpha_{\mu\nu}\hat{a}^{\dagger}_{\mu}\hat{a}_{\nu} +\sum_{pq}\alpha_{pq}\hat{a}^{\dagger}_{p}\hat{a}_{q} +\sum_{\mu p}(\alpha_{\mu p}\hat{a}^{\dagger}_{\mu}\hat{a}_{p}+\alpha^{*}_{\mu p}\hat{a}^{\dagger}_{p}\hat{a}_{\mu}) \nonumber \\
 & & +\sum_{pm}(\beta_{pm}\hat{a}^{\dagger}_{p}\hat{b}^{\dagger}_{m}+\beta^{*}_{pm}\hat{b}_{m}\hat{a}_{p}) +\sum_{\mu m}(\beta_{\mu m}\hat{a}^{\dagger}_{\mu}\hat{b}^{\dagger}_{m}+\beta^{*}_{\mu m}\hat{b}_{m}\hat{a}_{\mu}) +\sum_{mn}\gamma^{*}_{mn}\hat{b}_{n}\hat{b}^{\dagger}_{m} \biggr) \Biggr\}.
\end{eqnarray}
Hereafter, we use $\mu,\nu$ for electrons above the Fermi level, $p,q$ for electrons below the Fermi level, and $m,n$ for positrons, and $i,j,k,l$ denote arbitrary states. In the above equation, only the third and the fifth terms in the exponent give physical effects, and the others describe only the phase degree of freedom. Then we negrect the latter and
\begin{eqnarray}
\exp\Biggl\{i\biggl(\sum_{\mu p}(\alpha_{\mu p}\hat{a}^{\dagger}_{\mu}\hat{a}_{p}+\alpha^{*}_{\mu p}\hat{a}^{\dagger}_{p}\hat{a}_{\mu})+\sum_{\mu m}(\beta_{\mu m}\hat{a}^{\dagger}_{\mu}\hat{b}^{\dagger}_{m}+\beta^{*}_{\mu m}\hat{b}_{m}\hat{a}_{\mu})\biggr)\Biggr\}.
\end{eqnarray}
We will use this operator. Now we expand the density matrix in (82) by the parameters, and rearrange them in the order of the parameters. As the 0th order terms we obtain
\begin{eqnarray}
\sum_{ij}T^{\pm\pm}_{ij}(\rho^{++(0)}_{ji}+\rho^{--(0)}_{ji})+\frac{1}{2}\sum_{ijkl}(V^{\pm\pm\pm\pm}_{ijkl}-V^{\pm\pm\pm\pm}_{ilkj})\{\rho^{++(0)}_{ji}+\rho^{--(0)}_{ji}\}\{\rho^{++(0)}_{lk}+\rho^{--(0)}_{lk}\},
\end{eqnarray}
and it corresponds to $\langle \Phi_{0}|\hat{H}|\Phi_{0}\rangle$, the HF total energy, given in (39) and (40). Next the 1st order terms are given as
\begin{eqnarray}
 & & \sum_{ij}T^{\pm\pm}_{ij}\{\rho^{++(1)}_{ji}(\kappa)+\rho^{+-(1)}_{ji}(\kappa)+\rho^{-+(1)}_{ji}(\kappa)+\rho^{--(1)}_{ji}(\kappa)\} \nonumber \\
 & & +\sum_{ijkl}(V^{\pm\pm\pm\pm}_{ijkl}-V^{\pm\pm\pm\pm}_{ilkj}) \nonumber \\
 & & \times \{\rho^{++(1)}_{ji}(\kappa)+\rho^{+-(1)}_{ji}(\kappa)+\rho^{-+(1)}_{ji}(\kappa)+\rho^{--(1)}_{ji}(\kappa)\}\{\rho^{++(0)}_{lk}+\rho^{--(0)}_{lk}\}, \nonumber \\
 & & 
\end{eqnarray}
which correspond to $i\langle\Phi_{0}|[\hat{H},\hat{S}]|\Phi_{0}\rangle$. Calculate each matrix element and rearrange them, and we obtain 
\begin{eqnarray}
& & \sum_{\mu p}( T^{++}_{p\mu}i\alpha_{\mu p}-T^{++}_{\mu p}i\alpha^{*}_{\mu p})+\sum_{\mu n}( T^{-+}_{m\mu}i\beta_{\mu m}-T^{+-}_{\mu m}i\beta^{*}_{\mu m}) \nonumber \\
& & +\frac{1}{2}\biggl(\sum_{l<-m}+\sum_{-m<l\le F}-\sum_{l>F}\biggr)\Biggl\{ \sum_{\mu p}\Bigl((V^{++\pm\pm}_{p\mu ll}-V^{+\pm\pm+}_{pll\mu})i\alpha_{\mu p}-(V^{++\pm\pm}_{\mu pll} \nonumber \\
& & -V^{+\pm\pm+}_{\mu llp})i\alpha^{*}_{\mu p}\Bigr) + \sum_{\mu m}\Bigl((V^{-+\pm\pm}_{m\mu ll}-V^{-\pm\pm+}_{m ll\mu})i\beta_{\mu m} - (V^{+-\pm\pm}_{\mu mll}-V^{+\pm\pm-}_{\mu llm})i\beta^{*}_{\mu m}\Bigr) \Biggr\}.
\end{eqnarray}
Here if we take the derivative with respect to parameters, then we obtain the HF conditions (they are the same as the HF equations) and which coincide with previous definitions given by (41) and (42). The second order terms become
\begin{eqnarray}
& & \sum_{ij}T^{\pm\pm}_{ij}\{ \rho^{++(2)}_{ji}(\kappa)+\rho^{+-(2)}_{ji}(\kappa)+\rho^{-+(2)}_{ji}(\kappa)+\rho^{--(2)}_{ji}(\kappa) \} \nonumber \\
& & +\sum_{ijkl}(V^{\pm\pm\pm\pm}_{ijkl}-V^{\pm\pm\pm\pm}_{ilkj})\{(\rho^{++(0)}_{lk}+\rho^{--(0)}_{lk}) \nonumber \\
& & \times (\rho^{++(2)}_{ji}(\kappa)+\rho^{+-(2)}_{ji}(\kappa)+\rho^{-+(2)}_{ji}(\kappa)+\rho^{--(2)}_{ji}(\kappa)) \nonumber \\
& & +\frac{1}{2}(\rho^{++(1)}_{ji}(\kappa)+\rho^{+-(1)}_{ji}(\kappa)+\rho^{-+(1)}_{ji}(\kappa)+\rho^{--(1)}_{ji}(\kappa)) \nonumber \\
& & \times (\rho^{++(1)}_{lk}(\kappa)+\rho^{+-(1)}_{lk}(\kappa)+\rho^{-+(1)}_{lk}(\kappa)+\rho^{--(1)}_{lk}(\kappa)) \}.
\end{eqnarray}
It corresponds to $\frac{i^{2}}{2!}\langle \Phi_{0}|[[\hat{H},\hat{S}],\hat{S}]|\Phi_{0}\rangle $. We also calculate each terms, take into account the symmetries of matrix elements, Hermiticity of parameters, and choose the diagonal representation for the HF Hamiltonian, then we obtain 
\begin{eqnarray}
& & \sum_{\mu p}\Bigl( T^{++}_{\mu\mu}+\frac{1}{2}(\sum_{l<-m}+\sum_{-m<l \le F}-\sum_{l>F})(V^{++\pm\pm}_{\mu\mu ll}-V^{+\pm\pm+}_{\mu ll\mu}) \nonumber \\
& & -T^{++}_{pp}-\frac{1}{2}(\sum_{l<-m}+\sum_{-m<l \le F}-\sum_{l>F})(V^{++\pm\pm}_{ppll}-V^{+\pm\pm+}_{pllp})\Bigr)\alpha^{*}_{\mu p}\alpha_{\mu p} \nonumber \\
& & +\sum_{\nu m}\Bigl( T^{++}_{\nu\nu}+\frac{1}{2}(\sum_{l<-m}+\sum_{-m<l \le F}-\sum_{l>F})(V^{++\pm\pm}_{\nu\nu ll}-V^{+\pm\pm+}_{\nu ll \nu}) \nonumber \\
& & -T^{--}_{mm}-\frac{1}{2}(\sum_{l<-m}+\sum_{-m<l \le F}-\sum_{l>F})(V^{--\pm\pm}_{mmll}-V^{-\pm\pm-}_{mllm})\Bigr)\beta^{*}_{\nu m}\beta_{\nu m} \nonumber \\
& & +\sum_{\mu p}\sum_{\nu q} \Bigl\{ (V^{++++}_{\mu pq \nu}-V^{++++}_{\mu\nu qp})\alpha^{*}_{\mu p}\alpha_{\nu q} \nonumber \\
& & -\frac{1}{2}(V^{++++}_{p\mu q\nu}-V^{++++}_{p\nu q\mu})\alpha_{\mu p}\alpha_{\nu q}-\frac{1}{2}(V^{++++}_{\mu p\nu q}-V^{++++}_{\mu q \nu p})\alpha^{*}_{\mu p}\alpha^{*}_{\nu q} \Bigr\} \nonumber \\
& & + \sum_{\mu m}\sum_{\nu n}\Bigl\{(V^{-++-}_{m\mu\nu n}-V^{--++}_{mn\nu\mu})\beta^{*}_{\mu m}\beta_{\nu n} \nonumber \\
& & -\frac{1}{2}(V^{+-+-}_{\mu m\nu n}-V^{+-+-}_{\mu n\nu m})\beta^{*}_{\mu m}\beta^{*}_{\nu n}-\frac{1}{2}(V^{-+-+}_{m\mu n\nu}-V^{-+-+}_{m\nu n\mu})\beta_{\mu m}\beta_{\nu n} \Bigr\} \nonumber \\
& & +\sum_{\mu p}\sum_{\nu n}\Bigl\{(V^{+++-}_{p\mu\nu n}-V^{+-++}_{pn\nu\mu})\alpha_{\mu p}\beta^{*}_{\nu n} -(V^{+++-}_{\mu p \nu n}-V^{+-++}_{\mu n\nu p})\alpha^{*}_{\mu p}\beta^{*}_{\nu n} \nonumber \\
& & +(V^{++-+}_{\mu pn\nu}-V^{++-+}_{\mu\nu np})\alpha^{*}_{\mu p}\beta_{\nu n}-(V^{++-+}_{p\mu n\nu}-V^{++-+}_{p\nu n\mu})\alpha_{\mu p}\beta_{\nu n} \Bigr\}.
\end{eqnarray} 
Now the TDHF equations from which the RPA equations are derived become
\begin{eqnarray}
i\frac{\partial}{\partial t}\rho^{++}_{p\mu}(\kappa) &=& -\frac{\partial}{\partial i\alpha_{\mu p}}\langle \Phi(\kappa)|\hat{H}|\Phi(\kappa) \rangle, \\
i\frac{\partial}{\partial t}\rho^{++}_{\mu p}(\kappa) &=& -\frac{\partial}{\partial i\alpha^{*}_{\mu p}}\langle \Phi(\kappa)|\hat{H}|\Phi(\kappa) \rangle, \\
i\frac{\partial}{\partial t}\rho^{-+}_{m\nu}(\kappa) &=& -\frac{\partial}{\partial i\beta_{\nu m}}\langle \Phi(\kappa)|\hat{H}|\Phi(\kappa) \rangle, \\
i\frac{\partial}{\partial t}\rho^{+-}_{\nu m}(\kappa) &=& -\frac{\partial}{\partial i\beta^{*}_{\nu m}}\langle \Phi(\kappa)|\hat{H}|\Phi(\kappa) \rangle. 
\end{eqnarray}
Expand both sides with respect to the parameters as
\begin{eqnarray}
& & i\frac{\partial}{\partial t}\Bigl\{\rho^{++(0)}_{p\mu}+\rho^{++(1)}_{p\mu}(\kappa)+\rho^{++(2)}_{p\mu}(\kappa)+ \cdots \Bigr\}  \nonumber \\
& & = -\frac{\partial}{\partial i\alpha_{\mu p}}\biggl\{\langle \Phi_{0}|\hat{H}|\Phi_{0}\rangle + i\langle \Phi_{0}|[\hat{H},\hat{S}]|\Phi_{0}\rangle + \frac{i^{2}}{2!}\langle \Phi_{0}|[[\hat{H},\hat{S}],\hat{S}]|\Phi_{0}\rangle + \cdots \biggr\}, 
\end{eqnarray}
and so forth. In this expansion, the first term of each side is a constant, and the second term in the right-hand side becomes zero when HF condition is satisfied. Therefore, taking the lowest order non-zero matrix elements of both sides, we obtain
\begin{eqnarray}
i\frac{\partial}{\partial t}\rho^{++(1)}_{p\mu}(\kappa) &=& -\frac{\partial}{\partial i\alpha_{\mu p}}\frac{i^{2}}{2!}\langle \Phi_{0}|[[\hat{H},\hat{S}],\hat{S}]|\Phi_{0}\rangle, \\
i\frac{\partial}{\partial t}\rho^{++(1)}_{\mu p}(\kappa) &=& -\frac{\partial}{\partial i\alpha^{*}_{\mu p}}\frac{i^{2}}{2!}\langle \Phi_{0}|[[\hat{H},\hat{S}],\hat{S}]|\Phi_{0}\rangle, \\
i\frac{\partial}{\partial t}\rho^{-+(1)}_{m\nu}(\kappa) &=& -\frac{\partial}{\partial i\beta_{\nu m}}\frac{i^{2}}{2!}\langle \Phi_{0}|[[\hat{H},\hat{S}],\hat{S}]|\Phi_{0}\rangle, \\
i\frac{\partial}{\partial t}\rho^{+-(1)}_{\nu m}(\kappa) &=& -\frac{\partial}{\partial i\beta^{*}_{\nu m}}\frac{i^{2}}{2!} \langle \Phi_{0}|[[\hat{H},\hat{S}],\hat{S}]|\Phi_{0}\rangle. 
\end{eqnarray}
We assume that the system is in small normal-oscillations at a stationary point, and the number of the normal-oscillations equals the number of one-particle excitations. Then we can expand each freedom by the normal-oscillations
\begin{eqnarray}
\alpha_{\mu p} &=& \sum_{k}(P^{k}_{\mu p}e^{-i\omega_{k}t}+Q^{k}_{\mu p}e^{+i\omega_{k}t}), \\
\alpha^{*}_{\mu p} &=& \sum_{k}(P^{k*}_{\mu p}e^{+i\omega_{k}t}+Q^{k*}_{\mu p}e^{-i\omega_{k}t}), \\
\beta_{\nu m} &=& \sum_{k}(R^{k}_{\nu m}e^{-i\omega_{k}t}+S^{k}_{\nu m}e^{+i\omega_{k}t}), \\
\beta^{*}_{\nu m} &=& \sum_{k}(R^{k*}_{\nu m}e^{+i\omega_{k}t}+S^{k*}_{\nu m}e^{-i\omega_{k}t}).
\end{eqnarray}
Adopting them to the former relations we have derived, and picing up by the same frequency, we obtain the following relations (with complex conjugates) from (95)$\sim$(98)
\begin{eqnarray}
\omega_{k}P^{k}_{\mu p} &=& (\epsilon^{(+)}_{\mu}-\epsilon^{(+)}_{p})P^{k}_{\mu p} \nonumber \\
& & +\sum_{\nu q}\Bigl\{(V^{++++}_{\mu pq \nu}-V^{++++}_{\mu \nu qp})P^{k}_{\nu q}-(V^{++++}_{\mu p\nu q}-V^{++++}_{\mu q\nu p})Q^{k*}_{\nu q}\Bigr\} \nonumber \\
& & +\sum_{\nu n}\Bigl\{(V^{++-+}_{\mu pn \nu}-V^{++-+}_{\mu\nu np})R^{k}_{\nu n}-(V^{+++-}_{\mu p\nu n}-V^{+-++}_{\mu n\nu p})S^{k*}_{\nu n}\Bigr\}, \\
-\omega_{k}Q^{k}_{\mu p} &=& (\epsilon^{(+)}_{\mu}-\epsilon^{(+)}_{p})Q^{k}_{\mu p} \nonumber \\
& & +\sum_{\nu q}\Bigl\{(V^{++++}_{\mu pq\nu}-V^{++++}_{\mu\nu qp})Q^{k}_{\nu q}-(V^{++++}_{\mu p\nu q}-V^{++++}_{\mu q\nu p})P^{k*}_{\nu q}\Bigr\} \nonumber \\
& & +\sum_{\nu n}\Bigl\{(V^{++-+}_{\mu pn \nu}-V^{++-+}_{\mu\nu np})S^{k}_{\nu n}-(V^{+++-}_{\mu p\nu n}-V^{+-++}_{\mu n\nu p})R^{k*}_{\nu n} \Bigr\}, \\
-\omega_{k}S^{k*}_{\nu m} &=& (\epsilon^{(+)}_{\nu}+\epsilon^{(-)}_{m})S^{k*}_{\nu m} \nonumber \\
& & +\sum_{\mu n}\Bigl\{(V^{-++-}_{m\nu\mu n}-V^{--++}_{mn\mu\nu})S^{k*}_{\mu n}-(V^{-+-+}_{m\nu n\mu}-V^{-+-+}_{m\mu n\nu})R^{k}_{\mu n}\Bigr\} \nonumber \\
& & +\sum_{\mu p}\Bigl\{(V^{-+++}_{m\nu\mu p}-V^{-+++}_{mp\mu\nu})Q^{k*}_{\mu p}-(V^{-+++}_{m\nu p\mu}-V^{-+++}_{m\mu p\nu})P^{k}_{\mu p}\Bigr\}, \\
\omega_{k}R^{k*}_{\nu m} &=& (\epsilon^{(+)}_{\nu}+\epsilon^{(-)}_{m})R^{k*}_{\nu m} \nonumber \\
& & +\sum_{\mu n}\Bigl\{(V^{-++-}_{m\nu\mu n}-V^{--++}_{mn\mu\nu})R^{k*}_{\mu n}-(V^{-+-+}_{m\nu n\mu}-V^{-+-+}_{m\mu n\nu})S^{k}_{\mu n}\Bigr\} \nonumber \\
& & +\sum_{\mu p}\Bigl\{(V^{++-+}_{\mu pm\nu}-V^{++-+}_{\mu\nu mp})P^{k*}_{\mu p}-(V^{++-+}_{p\mu m\nu}-V^{++-+}_{p\nu m\mu})Q^{k}_{\mu p}\Bigr\}.
\end{eqnarray}
These are the general eigenvalue equations in our RPA theory. It is easy to obtain the polarization function 
\begin{eqnarray}
\Pi^{(0)}_{\mu\nu}({\bf x},{\bf y},\omega) &=& -i\int \frac{d\epsilon}{2\pi}{\rm tr}[\gamma_{\mu}S^{HF}_{F}({\bf x},{\bf y},\epsilon+\omega)\gamma_{\nu}S^{HF}_{F}({\bf y},{\bf x},\epsilon)],
\end{eqnarray}
from our RPA equations by use of a procedure employed in the nonrelativistic theory. These equations treat the Dirac vacuum on an equal footing with the $N$-electrons, as a many-body systems. 

Ionescu et al. also obtained an RPA equation for Dirac vacuum\cite{17}, by using the method of Bethe-Salpeter equation\cite{10}. They studied the collectivity of Dirac vacuum around heavy nucleus, but could not find the evidence of it.

\section{suggested applications}
\label{sec:app}

In this section, we propose some applications of our theory. Because it is besed on a simple and natural formulation, we should find many physical situations for our theory. Here we limit our discussion to the  atomic physical situations. 

In the context of the electronic structure of atoms, there are three effects in atoms: The electron correlation, the relativistic effect and the QED effect\cite{18}. 
The electronic structure of 
atoms is determined by the relation of these three factors. 
The electron correlation depends on the electron numbers and can be treated by several many-body methods. The relativistic effect becomes large with increasing the atomic number. This can be treated satisfactorily by the Dirac-Coulomb-Breit no-sea scheme. The case of highly ionized heavy atoms, the QED effect becomes dominant and it can not be neglected. This effect is treated by the perturbation theory in the Furry picture of QED. But practically this method can treat only the system of
atoms with a few electrons. In fact, especially in heavy atoms, as the ionicity
becomes high and the electron number decrease, the many-body effect becomes 
small and the QED effect becomes large. 
Therefore we argue that our theory should be applied to the cases where the electron number is not few and the many-body effects must be take into account at least in the mean-field level ( can not apply Furry picture perturbation scheme, in practice )
, also the QED effects are strong and can not be neglected. Thus heavy elements in middle level of ionicity should be one of the subject of our theory. The
Dirac-Coulom-Breit no-sea scheme has been applied to only neutral or almost neutral atoms. The method of the
QED correction works only for the highly ionized heavy atoms. 
Our theory should be suitable for intermediate case between them. 
We can make mention that only parts of the whole domain of the atomic physics 
was studied. 

To explain the excited states or optical phenomena of the system, 
we need the information about the many physical properties,
such as the excitation energies, transition amplitudes, 
multipole polarizabilities
and oscillator strengths\cite{19}. The RPA can treat these properties. 
The RPA method contains specific diagrams in all orders. 
Thus generally the RPA gives high numerical accuracy. 
The inner-core excitation or ionization in heavy elements like uranium, 
the QED effect is strong and we need a quantum electrodynamical theory. 
Our RPA ( Eqs. (103) $\sim$ (106) )should be applied. 
Our RPA is suitable for the phenomena where both the electron correlation and 
QED effect is important. 

In the heavy-ion collision, if the nuclear charges are large enough, 
a spontaneous electron-positron pair creation occurs. Our RPA can give
all types of excitations ( electron excitation, ionization, pair-creation ) which will be happened in the heavy-ion collision.
Futhermore, in these collisions, the adiabaticity breaks down. 
In such a strange chemical reaction we should treat the dynamics of the
electrons and nuclei on an equal footing. There is a method that does not use
the adiabatic potential energy sufaces but uses the time-dependent variational principle\cite{20}. Our TDHF ( Eqs. (77)$\sim$(80) ) can be combined with such a method to describe the heavy-ion collisions.
 
The problem of the renormalization still remains. But if it is solved
in the future, our method should gain a large ability for numerical
calculations, and should find various applications especially in the
atomic physics. We should obtain several maps or charts for various physical quantities in the atomic physics, which they have two coordinates ( atomic number and ionicity ).

\section{Concluding Remarks}
\label{sec:con}

In this paper, we have investigated a method to construct a quantum electrodynamical many-body theory, especially the mean-field theory. We have given the form of the HF in QED, and also some problems were discussed. We have taken the result obtained there as a starting point of our study, and have investigated the operator method. A relativistic Hamiltonian was derived on QED. The relativistic Slater determinant of Thouless form was obtained. We have discussed that in our case, the Wick theorem 
should be modified. After those preparations, we have introduced the TDHF and 
RPA in our theory. In our observation, the theory showed no difficulty at least
in the mean-field level. All of our theories include the Dirac sea as the dynamical degrees of freedom. This point is the characteristic feature of our formalism. This feature extends the capabilities of our theories than others. 

We should point out the possibility of the path-integral formalism. In a non-relativistic theory, 
Kuratsuji and Suzuki developed the method using generalized coherent state (GCS) representation\cite{21}. The crutial point of this method is using GCS for obtaining the resolution of unity and construct a kernel in path-integral form. They gave a proof that the Slater 
determinant of Thouless form can be used as GCS. We propose to apply this method to our theory.
In the imaginary time formalism, the partition function becomes 
\begin{eqnarray}
Z &=& {\rm Tr} e^{-\beta \hat{H}} \nonumber \\
  &=& \int {\cal D}(\alpha,\beta,\beta^{*},\gamma^{*}(\tau))\exp\Bigl[ -\int^{\beta}_{0} d\tau \langle \alpha,\beta,\beta^{*},\gamma^{*}(\tau)|\frac{\partial}{\partial \tau}+\hat{H}|\alpha,\beta,\beta^{*},\gamma^{*}(\tau) \rangle \Bigr], 
\end{eqnarray}
here $|\alpha,\beta,\beta^{*},\gamma^{*}\rangle$ is the Slater determinant and $\tau$ is the imaginary time. The path is determined by the movement of the coordinate ($\alpha, \beta, \beta^{*}, \gamma^{*}$). This kernel is exact for our Hamiltonian. If we apply the stationaly phase approximation, we obtain the time-dependent veriation principle leading to the TDHF equation, (63). Expand in the vicinity of the HF state by parameters, the following form is obtained: 
\begin{eqnarray}
Z &=& e^{-\beta E_{HF}}\int {\cal D}(\alpha,\beta,\beta^{*},\gamma^{*}(\tau))\exp\Bigl[ -\int^{\beta}_{0} d\tau \bigl\{  \langle \Phi_{0}|[\hat{S},\dot{\hat{S}}]|\Phi_{0}\rangle -\frac{1}{2}\langle\Phi_{0}|[[\hat{H},\hat{S}],\hat{S}]|\Phi_{0}\rangle \bigr\} \Bigr]. \nonumber \\
 & & 
\end{eqnarray}
 
About futher progress, we mention about the method of beyond-RPA. 
The RPA describes the small oscillations at a stationary point and 
only suitable for describing the low excitations.
We can investigate, in principle, the theory for high excitation and large amplitude anharmonic oscillations, such as the mode coupling theory. This theory should be applied to the system which is under both the strong QED effect and strong quantum fluctuation. We can also extend our theory to the non-equilibrium statistical mechanics through the TDHF formalism, the same way as the nonrelativistic theory. The works for these problems should become in the future.

\acknowledgments

The auther would like to express his gratitude to Professors. K. Higashijima and Y. Nambu, for many helpful discussions.

\appendix
\section*{The finite-temperature Hartree-Fock theory}
In this appendix, we derive the quantum electrodynamical finite-temperature Hartree-Fock theory by using the temperature Green's function formalism. To our knowledge, no finite-temperature HF theory has been presented in literature. But of course, there is the finite-temperature QED in literature\cite{22}. 

First, we give several definitions. The grand partition function and statistical operator are given as
\begin{eqnarray}
Z_{G} &=& e^{-\beta\Omega} = {\rm Tr} e^{-\beta \hat{K}}, \\
\hat{\rho}_{G} &=& Z^{-1}_{G}e^{-\beta\hat{K}} = e^{\beta(\Omega-\hat{K})}.
\end{eqnarray}
The Hamiltonian for the system is given as
\begin{eqnarray}
\hat{K} &=& \hat{K}_{0} + \hat{K}_{I}, \\ 
\hat{K}_{0} &=& \hat{H}_{0} -\mu \hat{N} = \int d^{3}x \hat{\bar{\psi}}({\bf x}\tau)\Bigl(-i\gamma\cdot\nabla + m_{0} + \gamma_{0}U(r) -\gamma_{0}\mu \Bigr)\hat{\psi}({\bf x}\tau), \\
\hat{K}_{I} &=& \hat{H}_{I} = -e\int d^{3}x\hat{\bar{\psi}}({\bf x}\tau)\gamma^{\mu}\hat{A}_{\mu}({\bf x}\tau)\hat{\psi}({\bf x}\tau). 
\end{eqnarray}
The field operators in the Heisenberg representation
\begin{eqnarray}
\hat{\psi}^{K}_{\alpha}({\bf x}\tau) &=& e^{\hat{K}\tau}\hat{\psi}_{\alpha}({\bf x})e^{-\hat{K}\tau}, \\
\hat{\bar{\psi}}^{K}_{\beta}({\bf x}\tau) &=& e^{\hat{K}\tau}\hat{\bar{\psi}_{\beta}}({\bf x})e^{-\hat{K}\tau}, \\
\hat{A}^{K}_{\mu}({\bf x}\tau) &=& e^{\hat{K}\tau}\hat{A}_{\mu}({\bf x})e^{-\hat{K}\tau}, 
\end{eqnarray}
where the first, the second and the third is the fermion, its Dirac conjugate and photon, respectively. The fermion field operators in the interaction representation are defined as
\begin{eqnarray}
\hat{\psi}^{I}({\bf x}\tau) &=& e^{\hat{K}_{0}\tau}\hat{\psi}({\bf x})e^{-\hat{K}_{0}\tau} \nonumber \\
 &=& \sum_{j} \Bigl( \psi^{(+)}_{j}({\bf x})e^{-(\epsilon^{(+)}_{j}-\mu)\tau}\hat{a}_{j} + \psi^{(-)}_{j}({\bf x})e^{+(\epsilon^{(-)}_{j}+\mu)\tau}\hat{b}^{\dagger}_{j} \Bigr),   \\
\hat{\bar{\psi}}^{I}({\bf x}\tau) &=& e^{\hat{K}_{0}\tau}\hat{\bar{\psi}}({\bf x})e^{-\hat{K}_{0}\tau} \nonumber \\
 &=& \sum_{j} \Bigl( \bar{\psi}^{(+)}_{j}({\bf x})e^{+(\epsilon^{(+)}_{j}-\mu)\tau}\hat{a}^{\dagger}_{j} + \bar{\psi}^{(-)}_{j}({\bf x})e^{-(\epsilon^{(-)}_{j}+\mu)\tau}\hat{b}_{j} \Bigr).
\end{eqnarray}
Then the free fermion temperature Green's function becomes 
\begin{eqnarray}
{\cal S}^{(0)}({\bf x}\tau,{\bf x'}\tau')_{\alpha\beta} &=& -\langle T_{\tau}[\hat{\psi}^{I}_{\alpha}({\bf x}\tau)\hat{\bar{\psi}}^{I}_{\beta}({\bf x'}\tau')]\rangle_{0} \nonumber \\
 &=& -\Bigl\{ \theta(\tau-\tau')\sum_{j}\Bigl(\psi^{(+)}_{j}({\bf x})\bar{\psi}^{(+)}_{j}({\bf x'})e^{-(\epsilon^{(+)}_{j}-\mu)(\tau-\tau')}(1-f(\epsilon^{(+)}_{j})) \nonumber \\
 & & + \psi^{(-)}_{j}({\bf x})\bar{\psi}^{(-)}_{j}({\bf x'})e^{+(\epsilon^{(-)}_{j}+\mu)(\tau-\tau')}f(\epsilon^{(-)}_{j})\Bigr)  \nonumber \\
 & & -\theta(\tau'-\tau)\sum_{j}\Bigl( \psi^{(+)}_{j}({\bf x})\bar{\psi}^{(+)}_{j}({\bf x'})e^{-(\epsilon^{(+)}_{j}-\mu)(\tau-\tau')}f(\epsilon^{(+)}_{j}) \nonumber \\
 & & \psi^{(-)}_{j}({\bf x})\bar{\psi}^{(-)}_{j}({\bf x'})e^{+(\epsilon^{(-)}_{j}+\mu)(\tau-\tau')}(1-f(\epsilon^{(-)}_{j})) \Bigr)\Bigr\}.
\end{eqnarray}
Here each fermion distribution function becomes
\begin{eqnarray}
f(\epsilon^{(+)}_{j}) = \frac{1}{e^{\beta(\epsilon^{(+)}_{j}-\mu)}+1}, \qquad f(\epsilon^{(-)}_{j}) = \frac{1}{e^{\beta(\epsilon^{(-)}_{j}+\mu)}+1}, 
\end{eqnarray}
and $\langle \hat{O}\rangle_{0}$ denotes the ensemble average $e^{\beta\Omega_{0}}{\rm Tr}\{e^{-\beta\hat{K}_{0}}\hat{O}\}$. Because of the fermion statistics, ${\cal S}$ satisfies the anti-periodic condition. We perform Fourier transformation in imaginary time. The Matsubara frequency will be odd, $\omega_{n}=(2n+1)\pi/ \beta$ and we obtain 
\begin{eqnarray}
{\cal S}^{(0)}({\bf x},{\bf x'},\omega_{n}) &=& \int^{\beta}_{0} d\tau e^{i\omega_{n}\tau}{\cal S}^{(0)}({\bf x},{\bf x'},\tau) \nonumber \\
 &=& -\sum_{j}\psi^{(+)}_{j}({\bf x})\bar{\psi}^{(+)}_{j}({\bf x'})\int^{\beta}_{0}d\tau e^{i\omega_{n}\tau}e^{-(\epsilon^{(+)}_{j}-\mu)\tau}(1-f(\epsilon^{(+)}_{j})) \nonumber \\
 & & -\sum_{j}\psi^{(-)}_{j}({\bf x)}\bar{\psi}^{(-)}_{j}({\bf x'})\int^{\beta}_{0}d\tau e^{i\omega_{n}\tau}e^{+(\epsilon^{(-)}_{j}+\mu)\tau}f(\epsilon^{(-)}_{j}) \nonumber \\
 &=& \sum_{j}\Biggl(\frac{\psi^{(+)}_{j}({\bf x})\bar{\psi}^{(+)}_{j}({\bf x'})}{i\omega_{n}-\epsilon^{(+)}_{j}+\mu}+\frac{\psi^{(-)}_{j}({\bf x})\bar{\psi}^{(-)}_{j}({\bf x'})}{i\omega_{n}+\epsilon^{(-)}_{j}+\mu}\Biggr).
\end{eqnarray}
The free photon temperature Green's function in the Feynman gauge is given by
\begin{eqnarray}
{\cal D}^{(0)}({\bf x}\tau,{\bf x'}\tau')_{\mu\nu} &=& -\langle T_{\tau}\hat{A}_{\mu}({\bf x}\tau)\hat{A}_{\nu}({\bf x'}\tau') \rangle_{0} \nonumber \\
 &=& \frac{1}{\beta}\sum_{n}\int\frac{d^{3}{\bf k}}{(2\pi)^{3}}e^{i{\bf k}\cdot({\bf x}-{\bf x'})-i\omega_{n}(\tau-\tau')}{\cal D}^{(0)}(\omega_{n},{\bf k})_{\mu\nu} \nonumber \\
 &=& \frac{1}{\beta}\sum_{n}\int\frac{d^{3}{\bf k}}{(2\pi)^{3}}e^{i{\bf k}\cdot({\bf x}-{\bf x'})-i\omega_{n}(\tau-\tau')}\frac{g_{\mu\nu}}{\omega_{n}^{2}+{\bf k}^{2}} \nonumber \\
 &=& \frac{1}{\beta}\sum_{n}g_{\mu\nu}\frac{e^{-|\omega_{n}||{\bf x}-{\bf x'}|}}{4\pi|{\bf x}-{\bf x'}|}e^{-i\omega_{n}(\tau-\tau')}, 
\end{eqnarray}
where ${\cal D}$ satisfieds the periodic condition because of the Bose statistics, and frequency becomes even $\omega_{n} = (2n)\pi/ \beta$.

Now we give the HF internal energy on a basis of the finite-temperature Feynman
rules and also of the results of previous sections. Take the symmetric form for the convergence factor, and we obtain
\begin{eqnarray}
E_{HF}(T,V,\mu) &=& \int d^{3}x \frac{1}{\beta} \sum_{n}\frac{1}{2}\Bigl(e^{i\omega_{n}\eta}\lim_{{\bf z}\to{\bf x}} + e^{-i\omega_{n}\eta}\lim_{{\bf z}\to{\bf x}} \Bigr) \nonumber \\
 & & \times (-i\gamma\cdot\nabla + m +\gamma_{0}U(r) -\gamma_{0}\mu ){\rm tr}{\cal S}^{HF}({\bf x},{\bf z},\omega_{n}) \nonumber \\
 & & +\frac{1}{2}e^{2}\int d^{3}x\int d^{3}y \frac{1}{\beta}\sum_{n}\frac{1}{\beta}\sum_{n'}{\cal D}^{(0)}({\bf x}-{\bf y},0)_{\mu\nu} \nonumber \\
 & & \times \frac{1}{2}\Bigl(e^{i\omega_{n}\eta}\lim_{{\bf z}\to{\bf x}}+e^{-i\omega_{n}\eta}\lim_{{\bf z}\to{\bf x}}\Bigr){\rm tr}\Bigl(\gamma^{\mu}{\cal S}^{HF}({\bf x},{\bf z},\omega_{n})\Bigr) \nonumber \\
 & & \times \frac{1}{2}\Bigl(e^{i\omega_{n'}\eta}\lim_{{\bf z'}\to{\bf y}}+e^{-i\omega_{n'}\eta}\lim_{{\bf z'}\to{\bf y}}\Bigr){\rm tr}\Bigl(\gamma^{\nu}{\cal S}^{HF}({\bf y},{\bf z'},\omega_{n'})\Bigr) \nonumber \\
 & & -\frac{1}{2}e^{2}\int d^{3}x \int d^{3}y \frac{1}{\beta}\sum_{n}\frac{1}{\beta}\sum_{n'}{\cal D}^{(0)}({\bf x}-{\bf y},\omega_{n}-\omega_{n'})_{\mu\nu} \nonumber \\
 & & \times {\rm tr}\Bigl\{ \gamma^{\mu}{\cal S}^{HF}({\bf x},{\bf y},\omega_{n})\frac{1}{2}(e^{i\omega_{n}\eta} + e^{-i\omega_{n}\eta}) \nonumber \\
 & & \times \gamma^{\nu}{\cal S}^{HF}({\bf y},{\bf x},\omega_{n'})\frac{1}{2}(e^{i\omega_{n'}\eta}+e^{-i\omega_{n'}\eta}) \Bigr\} \nonumber \\
 & & - E_{ZP},
\end{eqnarray}
where ${\cal S}^{HF}$ is the self-consistently determined HF temperature Green's function. Here the definition of zero-point energy is the same as (22). The internal energy depends temperature $T$ and chemical potential $\mu$. Perform summations for each frequency by the usual complex energy plane method. We obtain
\begin{eqnarray}
E_{HF}(T,V,\mu) &=& \frac{1}{2}\int d^{3}x \lim_{{\bf z}\to{\bf x}}(-i\gamma\cdot\nabla + m + \gamma_{0}U(r) -\gamma_{0}\mu ) \nonumber \\
 & & \times {\rm tr}\Bigl( \sum_{j}\psi^{(+)}_{j}({\bf x})\bar{\psi}^{(+)}_{j}({\bf z})(2f(\epsilon^{(+)}_{j})-1) + \sum_{j}\psi^{(-)}_{j}({\bf x})\bar{\psi}^{(-)}_{j}({\bf z})(1-2f(\epsilon^{(-)}_{j}))\Bigr) \nonumber \\
 & & +\frac{1}{2}e^{2}\int d^{3}x\int d^{3}y \frac{g_{\mu\nu}}{4\pi|{\bf x}-{\bf y}|} \nonumber \\
 & & \times \frac{1}{2}{\rm tr}\biggl\{ \gamma^{\mu}\Bigl(\sum_{i} \psi^{(+)}_{i}({\bf x})\bar{\psi}^{(+)}_{i}({\bf x})(2f(\epsilon^{(+)}_{i})-1) + \sum_{i}\psi^{(-)}_{i}({\bf x})\bar{\psi}^{(-)}_{i}({\bf x})(1-2f(\epsilon^{(-)}_{i}))\Bigr)\biggr\}  \nonumber \\
 & & \times \frac{1}{2}{\rm tr}\biggl\{ \gamma^{\nu}\Bigl(\sum_{j} \psi^{(+)}_{j}({\bf y})\bar{\psi}^{(+)}_{j}({\bf y})(2f(\epsilon^{(+)}_{j})-1) + \sum_{j}\psi^{(-)}_{j}({\bf y})\bar{\psi}^{(-)}_{j}({\bf y})(1-2f(\epsilon^{(-)}_{j}))\Bigr) \biggr\}  \nonumber \\
 & & -\frac{1}{2}e^{2}\int d^{3}x \int d^{3}y g_{\mu\nu}\frac{\exp(i|\epsilon^{(\pm)}_{i}-\epsilon^{(\pm)}_{j}||{\bf x}-{\bf y}|)}{4\pi|{\bf x}-{\bf y}|}\frac{1}{4}{\rm tr}\biggl\{  \nonumber \\
 & & \times \gamma^{\mu} \Bigl( \sum_{i}\psi^{(+)}_{i}({\bf x})\bar{\psi}^{(+)}_{i}({\bf y})(2f(\epsilon^{(+)}_{i})-1) + \sum_{i}\psi^{(-)}_{i}({\bf x})\bar{\psi}^{(-)}_{i}({\bf y})(1-2f(\epsilon^{(-)}_{i})) \Bigr) \nonumber \\
 & & \times \gamma^{\nu} \Bigl( \sum_{j}\psi^{(+)}_{j}({\bf y})\bar{\psi}^{(+)}_{j}({\bf x})(2f(\epsilon^{(+)}_{j})-1) + \sum_{j}\psi^{(-)}_{j}({\bf y})\bar{\psi}^{(-)}_{j}({\bf x})(1-2f(\epsilon^{(-)}_{j})) \Bigr) \biggr\} \nonumber \\
 & & -E_{ZP}.
\end{eqnarray}
The above equation includes the effect of the Dirac vacuum correctly. Take the zero-temperature limit, it recover the zero-temperature theory given in the previous sections. The Dirac vacuum also has thermal energy, depending on the temperature and be treated in the same footing as many-particles. (A16) also has the renormalization problem. Our result given in this appendix should be a starting point to the next step, for example, the operator formalism for finite-temperature mean-field dynamics.

\end{document}